\documentclass[twocolumn]{aastex62}

\begin{document}

\title{Star Forming Galaxies as AGN Imposters? A Theoretical Investigation of the Mid-infrared Colors of AGNs and Extreme Starbursts}

\author[0000-0003-2277-2354]{Shobita Satyapal}
\affiliation{George Mason University, Department of Physics \& Astronomy, MS 3F3, 4400 University Drive, Fairfax, VA 22030, USA}

\author[0000-0002-0785-9139]{Nicholas P. Abel}
\affiliation{MCGP Department, University of Cincinnati, Clermont College, Batavia, OH 45103, USA}

\author[0000-0002-4902-8077]{Nathan J. Secrest}
\affiliation{U.S.\ Naval Observatory, 3450 Massachusetts Avenue NW, Washington, DC 20392, USA}
\affiliation{George Mason University, Department of Physics \& Astronomy, MS 3F3, 4400 University Drive, Fairfax, VA 22030, USA}

\correspondingauthor{Shobita Satyapal}
\email{ssatyapa@gmu.edu}

\begin{abstract}

We conduct for the first time a theoretical investigation of the mid-infrared spectral energy distribution (SED) produced by dust heated by an active galactic nucleus (AGN) and an extreme starburst. These models employ an integrated modeling approach using photoionization and stellar population synthesis models in which both the line and emergent continuum is predicted from gas exposed to the ionizing radiation from a young starburst and an AGN. In this work, we focus on the infrared colors from the {\it Wide-field Infrared Survey Explorer}, predicting the dependence of the colors on the input radiation field, the ISM conditions, the obscuring column, and the metallicity. We find that an extreme starburst can mimic an AGN in two band mid-infrared color cuts employed in the literature. However, the three band color cuts employed in the literature require starbursts with extremely high ionization parameters or gas densities . We show that the extreme mid-IR colors seen in some blue compact dwarf galaxies are not due to metallicity but rather a combination of high ionization parameters and high column densities. 
Based on our theoretical calculations, we present a theoretical mid-infrared color cut that will exclude even the most extreme starburst that we have modeled in this work. The theoretical AGN demarcation region presented here can be used to identify elusive AGN candidates for future follow-up studies with the {\it James Webb Space Telescope (JWST)}. The full suite of simulated SEDs are available online.
\end{abstract}


\section{Introduction}
\label{intro}

Supermassive black holes (SMBHs), which manifest as active galactic nuclei (AGNs) when accreting, are now known to be a fundamental component of galaxies and play an important role in their evolution.  Detecting complete samples of AGNs unaffected by obscuration, and understanding their connection to the properties of the host galaxies in which they reside has therefore been an important goal.  Over the past several decades, it has become clear that a large fraction of AGNs are missed in optical surveys due either to obscuration of the central engine, or contamination of the optical emission lines from star formation in the host galaxy \citep[e.g.,][]{goulding2009,trump2015}. This is a significant deficiency, because these elusive AGNs are often found in key phases of galaxy evolution, such as late stage galaxy mergers, when the black hole is expected to grow most rapidly \citep{vanwassenhove2012,blecha2013}, or in low mass and bulgeless galaxies, a galaxy population that may place important constraints on models of SMBH seed formation and merger-free models to SMBH growth \citep[e.g.,][]{Volonteri2009, Volonteri2010, vanwassenhove2010, greene2012}.

Over the past decade, mid-infrared color selection has been shown to be a powerful tool in uncovering optically hidden AGNs in a large population of galaxies. This is because the hard radiation field associated with AGNs can heat the dust to temperatures as high as the grain sublimation temperature, producing a strong mid-infrared continuum and an infrared spectral energy distribution (SED) that is clearly distinguishable from typical star forming galaxies that is independent of obscuration of the central engine.  The all-sky survey carried out by the {\it Wide-field Infrared Survey Explorer} \citep[WISE;][]{wright2010} has opened up an unprecendented window in the search for optically hidden AGNs in a large number of galaxies, demonstrating that mid-infrared color selection can identify luminous AGNs with a reliability greater than 95\%  \citep[e.g.,][]{stern2012,mateos2015,secrest2015}. 

While the use of mid-infrared color selection in identifying powerful AGNs in which the AGN dominates over the stellar emission in the host galaxy is now on firm empirical ground, it is well-known that this method fails in galaxies in which the luminosity of the stellar emission from the host galaxy is comparable to the luminosity of the AGN. Indeed, only a small percentage ($< $10\%) of galaxies spectroscopically classified as AGNs from the Sloan Digital Sky Survey Data Release 7 (SDSS DR7)\footnote{\url{http://wwwmpa.mpa-garching.mpg.de/SDSS/DR7/}} are classified as AGNs based on the three-band mid-infrared color selection criterion from \citet{jarrett2011} (see \citet{yan2013}).  In addition, because the current suite of mid-infrared classification schemes are empirical and based on the location in mid-infrared color space of {\it known} samples of optically identified quasars or AGNs identified through hard X-ray surveys \citep[e.g.,][]{lacy2004,stern2005,donley2012,stern2012,assef2013,mateos2012}, there is no way to ensure that rare objects characterized by extreme star formation may be erroneously classified as AGNs using these schemes. Likewise, the stellar templates that are currently available are also empirical \citep[e.g.,][]{assef2010}, and therefore cannot be used to identify rare objects, or gain physical insight into the ISM conditions that give rise to the observed mid-infrared SED observed from a galaxy.

In recent years, there have been a number of studies that have revealed an unexpected population of optically classified star forming galaxies with extremely red mid-infrared colors suggestive of AGN activity  \citep[e.g.,][]{griffith2011,izotov2011,satyapal2014,sartori2015,secrestb2015,satyapal2016,oconnor2016,hainline2016}.   In many cases, these galaxies lack significant bulges or are low mass galaxies, a population that thus far  seemed to be quiescent based on multiwavelength studies. While the red mid-IR colors are highly suggestive of accretion activity, and follow-up multiwavelength studies provide some compelling support for this scenario in a few cases \citep[e.g.][]{secrestb2015,satyapal2016}, it is possible that the dust can be heated by star formation alone in the majority of cases.  Since  there are a number of blue compact dwarfs (BCDs) with extremely red mid-infrared colors \citep[e.g.,][]{griffith2011,izotov2011}, it has been suggested that star formation in extreme and perhaps low metallicity environments can produce an infrared spectral energy distribution similar to an AGN \citep[e.g.][]{satyapal2014,hainline2016}. 

In this work, we investigate the possibility that purely star forming galaxies can produce similar mid-infrared colors to AGNs while simultaneously producing the typical optical emission line ratios characteristic of observed star forming galaxies. The goal of the paper is to determine, based on theoretical SEDs, if an extreme starburst can mimic an AGN in its infrared colors. Unlike empirical stellar templates, these theoretical models can be used to gain physical insight into the relationship between the mid-infrared SED and the ionizing radiation field, and the properties of the surrounding ISM. In Section 2 we discuss the details of our theoretical models, stating all of our assumptions on the input radiation field, the metallicity, geometry, total column density, and gas properties. In Section 3, we discuss our main results, focussing on the mid-infrared colors predicted in the WISE bands as a function of our model parameters. In Section 4, we present a theoretical mid-infrared color-cut that can be employed to identify promising AGN candidates for follow-up study. In Section 5, we summarize our results.

\section {Theoretical Calculations}

In order to determine if a starburst incident radiation field is capable of producing dust emission features usually associated with AGN, while simultaneously reproducing the classical emission-line ratios of starbursts in the optical, we have performed a series of calculations using the spectral synthesis code Cloudy \citep{ferland2013}.  The ability of Cloudy to both model the emission-line spectrum along with a detailed model of dust \citep{vanHoof2004} makes it an ideal computational tool to study the sensitivity of the gas and dust spectrum to the input ionizing radiation field and the physical conditions of the gas.  In addition, Cloudy has the ability to compute the spectrum of multiple gas phases, including H II regions, photodissociation regions (PDRs), and molecular clouds, in a single calculation, by assuming an equation of state between each gas phase (usually constant density or constant pressure).  A complete list of physical processes treated by Cloudy in each physical regime can be found in \citet{ferland2013}. The shape of the infrared dust continuum and the emission line spectrum depends on several factors, including the shape of the ionizing radiation field, the geometrical distribution of the gas and the stopping criterion, the chemical composition and grain properties, the gas properties and the equation of state. Below we discuss the details of our model parameters. The full set of simulated SEDs are available to download online.\footnote{\url{http://physics.gmu.edu/~satyapal/syntheticspectra}} We note that we have not included the effects of shocks in our calculations.  Starburst driven winds can produce shock fronts that ionize the gas and alter the grain size distribution, possibly destroying small grains. This in turn will affect the heating of the grains and the resulting transmitted mid-infrared SED. However shocks will also affect the emission line spectrum and the galaxy's location on the BPT diagram \citep{allen2008,kewley2013}. Our study specifically determines if galaxies with an optical emission line spectrum typical of star forming galaxies can masquerade as AGNs based on their mid-infrared SED.

\subsection{Incident Radiation Field}
 
The emission line spectrum and the temperature of the dust and therefore the shape of the infrared continuum will depend on the spectral energy distribution (SED) of the ionizing radiation field incident on the gas. Our calculations are designed to model the emergent line and continuum emission from gas illuminated by  a mix of a starburst and AGN SED.  For the AGN, we used the SED given in \citet{korista1997} in their study of broad line region (BLR) line intensities. This continuum combines an X-ray power law model with a `blue bump' component of the form:
\begin{equation}
f_\nu=\nu^{\alpha_\mathrm{UV}}\exp\left(\frac{-h\nu}{kT_\mathrm{BB}}\right)\exp\left(\frac{-kT_\mathrm{IR}}{h\nu}\right)+a\nu^{\alpha_\mathrm{X}},
\end{equation}
\noindent where $T_\mathrm{BB}$ is the temperature of the ``Big Bump'',$\alpha_{ox}$ is the ratio of X-ray to UV flux, $\alpha_\mathrm{UV}$ is the low-energy slope of the Big Bump continuum, and $\alpha_\mathrm{X}$ is the slope of the X-ray continuum. The constant $a$ is adjusted to produce the correct UV to X-ray slope and $kT_\mathrm{IR}$ sets the infrared cutoff temperature of the Big Bump to 0.01 Rydbergs. For our calculations, we use T = $10^{6}$ K,   $\alpha_{ox} = -1.4$, $\alpha_\mathrm{UV} = -0.5$, and $\alpha_\mathrm{X} =-1.0$. 

For the starburst input SED, we used a 5 Myr continuous star-formation starburst model  from the Starburst 99 website\footnote{\url{http://www.stsci.edu/science/starburst99/docs/parameters.html}} \citep{leitherer99,Leitherer14} as our input SED with a Saltpeter initial mass function (IMF) of power-law index $\alpha=-2.35$ and a star-formation rate of 1 ~$M_\sun$~yr$^{-1}$, with lower and upper mass cutoffs of 1 and 100 $M_\sun$, respectively. All other parameters were left at their default settings from the Starburst99 website. Note that the default settings employ the Geneva stellar tracks and an''evolution'' wind model.  Since the primary goal of this work is to determine if a pure starburst can mimic an AGN in its mid-infrared continuum properties, our choice of starburst SED was governed by requiring the most extreme ionizing radiation field from a stellar population.  The emergence of Wolf-Rayet stars at 5 Myr, along with massive O stars results in the hardest radiation field from a starburst \citep{schaerer1996,kehrig2008}.  At these ages, the population of high mass X-ray binaries is also expected to peak (see Figure 1 in \citet{linden2010}), resulting in the highest X-ray luminosities produced by the starburst.We note that the upper mass cutoff of the IMF or the IMF slope has an insignificant effect on the ionizing flux and therefore the dust SED unless the upper mass cutoff falls below 30 $M_\sun$. This is because there are insufficient stars at the very high masses to contribute significantly to the integrated light of the galaxy.  When the upper mass cutoff falls below  30 $M_\sun$, however, there is a significant drop in the ionizing photon flux, resulting in a softer SED. Since this will create cooler dust, we consider only the most extreme starburst model in this work.

The total luminosity is fixed at $L=10^{43}$~erg~s$^{-1}$.  Note that the total luminosity and star formation rate do not affect the shape of the ionizing radiation field or the shape of the emergent SED for our assumed  extreme starburst model. We allow the fraction of the total luminosity that is due to AGN and starburst activity to vary, in increments of 20\%, such that the total luminosity is fixed but the relative contribution of each SED varies.  This leads to a total of six SED combinations (0\% AGN - 100\% AGN).  Most of our results will be involving the 100\% starburst calculations, which henceforth will be referred to as the ``pure starburst'' calculation.  Cosmic rays and the Cosmic Microwave Background (CMB) are both included, with the cosmic ray ionization rate set to $3~\times10^{-16}$~s$^{-1}$ \citep{indriolo2012}. We also study the effects of metallicity on the spectrum by selecting two separate metallicities. The first set of models assume solar metallicity, both for the stellar population as well as the interstellar medium (ISM).  For the lower metallicity models we use a different SED from Starburst99, reducing the metallicity to 0.10 solar, keeping the age the same as before, and reducing the gas metallicity by the same amount. Note that we self-consistently vary both the metallicity of the stellar population as well as the metallicity of the gas in our models. For the AGN models, only the ISM abundance is changed between the high and low metallicity models, where the abundance set and depletion factors are matched with the starburst models. It is possible that the varying the metallicity may affect the torus and accretion disk structure surrounding the black hole, altering the radiation field incident on the narrow line region. However, since the effect of metallicity on the intrinsic AGN SED is unknown, we assume the same incident ionizing radiation field for all AGN models presented in this work.

\par

\begin{figure}
\noindent{\includegraphics[width=8.7cm]{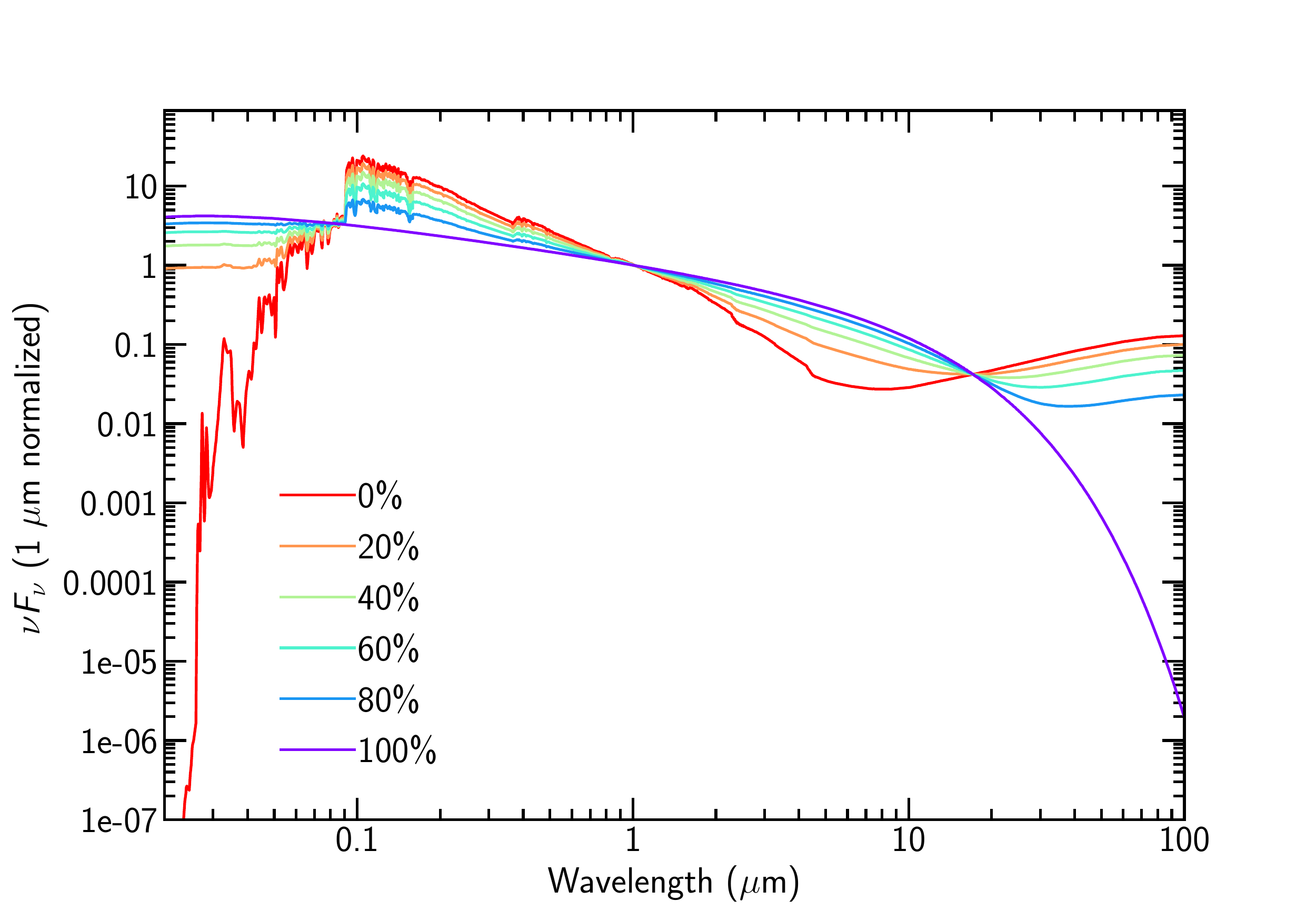}}
\caption{Incident composite SED for the solar metallicity starburst model with varying contributions of the AGN SED, ranging from 0\%-100\%, as indicated in the legend.}
\label{solarSED}
\end{figure}

\begin{figure}
\noindent{\includegraphics[width=8.7cm]{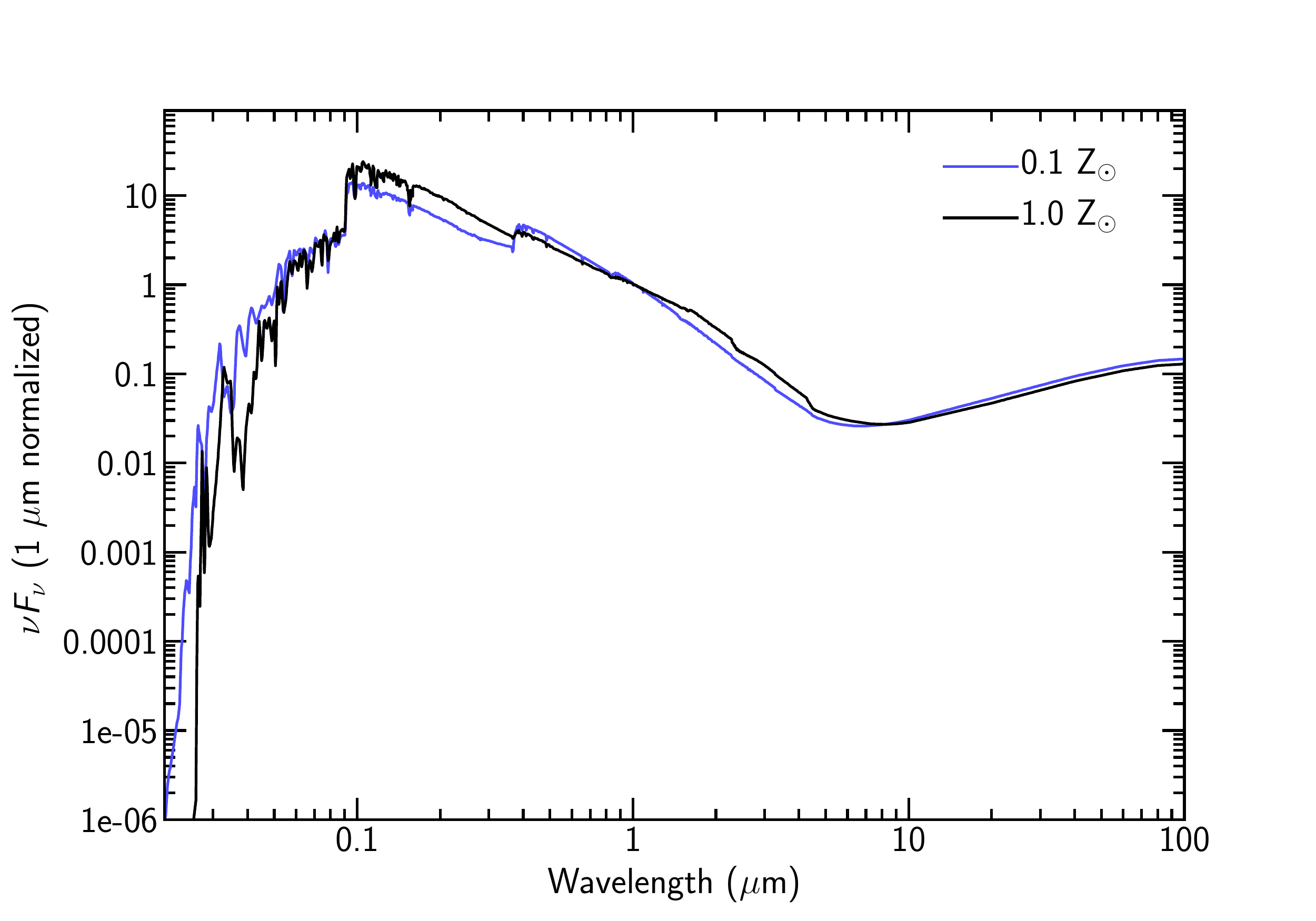}}
\caption{Comparison of the solar metallicity and low metallicity starburst models for a 5~Myr starburst.}
\label{metallicitycompareSED}
\end{figure}

In Figure~\ref{solarSED}, we plot the incident continuum for the full set of solar metallicity models.  As can be seen, the AGN SED extends to much higher energies than does the pure starburst model. The pure starburst model shows a prominent dip between the peak of the stellar emission and the infrared continuum in contrast to the pure AGN model, which displays an enhanced mid-infrared continuum due to hot dust emission from the torus, demonstrating the power of mid-infrared color selection in identifying energetically dominant AGNs (see \citet{donley2012}). In Figure~\ref{metallicitycompareSED}, we plot the SED for a 5 Myr starburst for the solar and low metallicity models. The metallicity affects the shape of the SED of the stellar population since the effective temperature of massive stars are higher at low metallicity, resulting in a harder spectrum.  In addition, metallicity has a significant effect on the stellar evolution since the lower mass loss rates in low metallicity stars result in longer main-sequence lifetimes of the most massive stars and the suppression of the formation and lifetime of the W-R phase \citep{levesque2006,meynet1994}.The net result of these various effects is a harder SED that extends to shorter wavelengths in the low metallicity model.

  \par
\subsection{Geometrical Distribution of the Gas}

The geometry of our calculation is a one-dimensional spherical model with a closed geometry, where the cloud is between the observer and the continuum source.  This geometry means, for thicker models, the optical spectrum can be extinguished along the line of sight.  We vary the inner radius of the cloud, $R$, from the center (where we assume both the starburst and the AGN is located) from $\log({R/\textrm{cm}})=19.29$ to 22.29.  The combination of inner radius, luminosity, and SED sets the total hydrogen ionizing flux striking the cloud and hence the ionization parameter, where the dimensionless ionization parameter, $U$, is defined as the dimensionless ratio of the incident ionizing photon density to the hydrogen density:
\begin{equation}
U=\frac{\phi_\mathrm{H}}{n_\mathrm{H}}=\frac{Q(\mathrm{H})}{4\pi R^{2}n_\mathrm{H}c},
\end{equation}

\noindent where $\phi_\mathrm{H}$ is the flux of hydrogen ionizing photons striking
the illuminated face of the cloud per second, $n_\mathrm{H}$ is the hydrogen density, $c$ is the speed
of light, and $Q(\mathrm{H})$ is the number of hydrogen ionizing photons
striking the illuminated face per second. By setting the density at the illuminated face (see below), $U$ is allowed to vary in our calculations from about $10^{-5.5}$ to $10^{1}$.  This covers the a large dynamic range of ionization parameters, which allows us to explore the theoretical effects of extreme ionization parameters on the infrared continuum. Note however that on average, $U$ is $\sim10^{-3}$ based on observations of the optical emission lines in star forming galaxies and HII regions \citep{dopita2000, moustakas2010}, but can be higher in regions such as ULIRGs \citep{abel2009} or high redshift galaxies \citep{pettini2001,brinchmann2008,maiolino2008,hainline2009,erb2010}. In general, models with a similar effective ionization parameter produce very similar emission line ratios, assuming all other parameters are held constant.  
\par

Our calculations end for a range of total hydrogen column densities $N_\mathrm{H}$, so we can study the effects of extinction on the dust and optical SED. The stopping criterion (corresponding to a particular $A_V$) is an important parameter in our model results. The thicker we make the cloud,
the cooler the dust for a given ionizing radiation field, which will affect the shape of the mid-infrared continuum, the depth of the 9.7~\micron\ silicate feature, and the strength of the PAH features since the abundance of PAHs varies with cloud depth (see section ~\ref{gasdust}). Additionally, the thicker the cloud, the larger the optical depths,
which will affect the overall line to continuum ratios emerging from the cloud. We allow $N_\mathrm{H}$ to vary from $10^{19}$ cm$^{-2}$ to $10^{24}$ cm$^{-2}$, in increments of 0.5 dex.  The upper limit of the column density considered by our calculations assure that, for a given U, the calculations extend beyond the hydrogen ionization front and into the PDR/molecular cloud. 

\par  

\subsection{Gas and Dust Properties}
\label{gasdust}

We chose the gas and dust abundances in our calculations to be consistent with the local ISM.  Our choice of gas phase abundances are identical to \citet{abel2009} and point the reader to this work for a complete list of abundances.  We adopt an $R_V = 3.1$ grain size distribution consistent with the ISM, where graphite and silicate grains are considered \citep[]{mathis1977, weingartner2001}.  Each grain constituent is divided into ten bins, with upper and lower limits to the grain sizes designed to reproduce the ISM extinction curve.  The gas to dust ratio, $A_V/N_\mathrm{H}$, used in our calculations is set to $5.4~\times10^{-22}$ mag cm$^{2}$.  The charges and temperatures of the grains are determined self-consistently.  We also include size-resolved polycyclic aromatic hydrocarbons
(PAHs) in our calculations, with the same size distribution
used by \citet{abel2008}. Note that since PAHs are thought to be destroyed from hydrogen ionizing radiation and coagulate in molecular environments (see for instance, \citet{omont1986} and \citet{Verstraete1996}), we scale the PAH abundance by $n^0_\mathrm{H}/N_\mathrm{H}$. The full physical model of dust employed in Cloudy is described in detail in \citet{vanHoof2004}, so we point the reader to this work for further details.  One caveat to both the gas and dust abundances is that, for the low metallicity models, we scale all metals along with the dust abundance by a factor of 0.10.  

\par
\subsection{Physical Conditions of the Gas and the Equation of state}

The simulated spectrum, which is computed beginning at the inner cloud radius, is affected by our choice for the equation of state, which determines the density and pressure with depth into the cloud.   The gas density affects the emission line ratios due to the dependence of the collisional excitation rates on electron density.  Also, for a given ionization parameter and incident SED, the temperature of the dust will be greater for smaller inner radii, $R$ (see  Equation 2), thereby changing the shape of the infrared continuum.   For each SED, radius $R$, and stopping depth, $N_\mathrm{H}$, we therefore consider two densities ($n_\mathrm{H}$) at the illuminated face and two equations of state. The initial density at the inner radius is set to 300~cm$^{-3}$, typical of local HII regions \citep{osterbrock1989}.  We also consider a more extreme density of $10^{3}$ cm$^{-3}$ to explore whether the most extreme  starbursts with the most extreme gas conditions could in principle mimic AGNs in their mid-infrared continuum properties. We note that in local galaxies, electron densities up to $\approx10^{3}$ cm$^{-3}$ are found in highly active star forming galaxies \citep{kewley2001b,armus2004}, but such high densities are rare in local galaxies. However, such conditions may be more prevalent in high redshift galaxies \citep{brinchmann2008,rigby2011}.

\par

For each initial density, we calculate the emitted gas and dust spectrum with depth for the case where the cloud is isobaric and for constant density.  We include thermal, radiation, and turbulent pressure in our models with the turbulent velocity set to 5.0 km s$^{-1}$ (see \citet{pellegrini2007} for an application of isobaric models to the study of M17, and \citet{abel2006} and \citet{abel2009}, for an application of constant pressure models to ULIRGs).  The assumption of constant pressure means the density will increase with depth into the cloud \citep{pellegrini2007} as the calculation reaches the hydrogen ionization front and moves into the PDR.  Each calculation is computed for two iterations, which is required for optically thick models in order to compute the emergent spectrum. 
\par

Given the full suite of composite SEDs, $n_\mathrm{H}$, $N_\mathrm{H}$, radius, metallicity, and equation of state, we computed a total of 8,184 simulations. For each model, we calculated the magnitudes in the WISE $W1$ (3.4 $\mu$m), $W2$ (4.6 $\mu$m), and  $W3$ (12 $\mu$m) bands using the WISE filter transmission curves.\footnote{http://wise2.ipac.caltech.edu/docs/release/prelim/expsup/sec4\_3g.html} In this work, we explore the range of parameter space in which the models can reproduce the three-band mid-infrared color space ($W1-W2$ versus $W2-W3$) from \citet{jarrett2011} widely used to separate AGNs from star forming galaxies in large surveys.   We note that while we focus on the mid-infrared colors from the WISE bands, our SEDs can be used to calculate fluxes synthesize colors in wavelength bands from any past or future mission, or for general SED fitting across a broad wavelength range. In a future paper, we explore the emission line spectrum predicted by these simulations.

\section {Results}

\subsection {The Dependence of the Infrared Continuum on Model Parameters}

\begin{figure}
\noindent{\includegraphics[width=8.7cm]{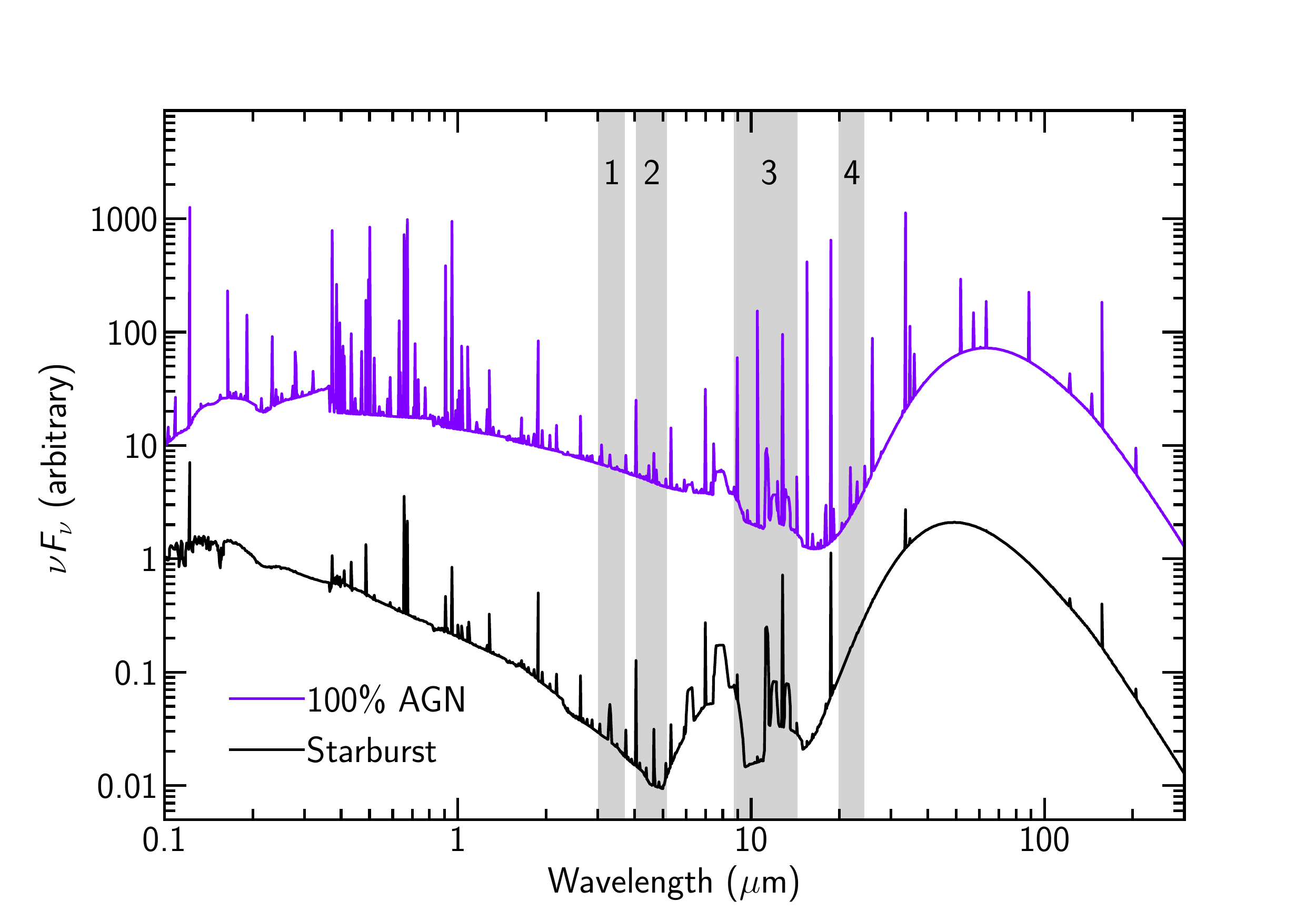}}
\caption{The transmitted continuum at a stopping column density of  $\log(N_\mathrm{H}/\textrm{cm}^{-2})=21$ for the pure starburst and the 100\% AGN SED for the solar metallicity and standard gas density of $n_\mathrm{H}=300$~cm$^{-3}$ simulation. Here we show a representative spectrum corresponding to an ionization parameter $\log(U)=-3.0$. The AGN SED results in hotter dust and weaker PAH emission compared with the starburst model. Note that for the AGN SED, the inner radius of the cloud is larger compared to the starburst model in order to obtain the same ionization parameter, lowering the dust temperature of the AGN model. Despite this effect, the AGN SED still produces an excess in the mid-infrared emission due to hotter dust compared with the starburst model at a given ionization parameter. The shaded region denotes the wavelength range of the first four WISE bands. The SEDs are offset by a factor of 10 for clarity.}
\label{transmitted_SED_starburstAGN}
\end{figure}

The shape of the mid-infrared continuum is strongly dependent on our model parameters. In Figure ~\ref{transmitted_SED_starburstAGN}, we plot the transmitted continuum at a stopping column density of $\log(N_\mathrm{H}/\textrm{cm}^{-2})=21$ for the pure starburst model and the 100\% AGN model for solar metallicity and a standard gas density of $n_\mathrm{H}=300$~cm$^{-3}$.  As can be seen, the shape of the infrared continuum depends strongly on the input SED, which directly affects the temperature of the dust. The harder radiation field from the AGN results in hotter dust and an excess in the continuum emission within the WISE bands. Note the strength of the PAH features is stronger in the starburst, since the harder radiation field from the AGN destroys the grains out to a larger column density. The spectrum of the AGN is also dominated by fine structure lines from highly ionized species with ionization potentials. These lines are absent in the pure starburst model. The analysis of the spectral signatures of an extreme starburst and AGN from our models will be presented in a future paper (Satyapal et al. 2018, in preparation).

\par

\begin{figure}
\noindent{\includegraphics[width=8.7cm]{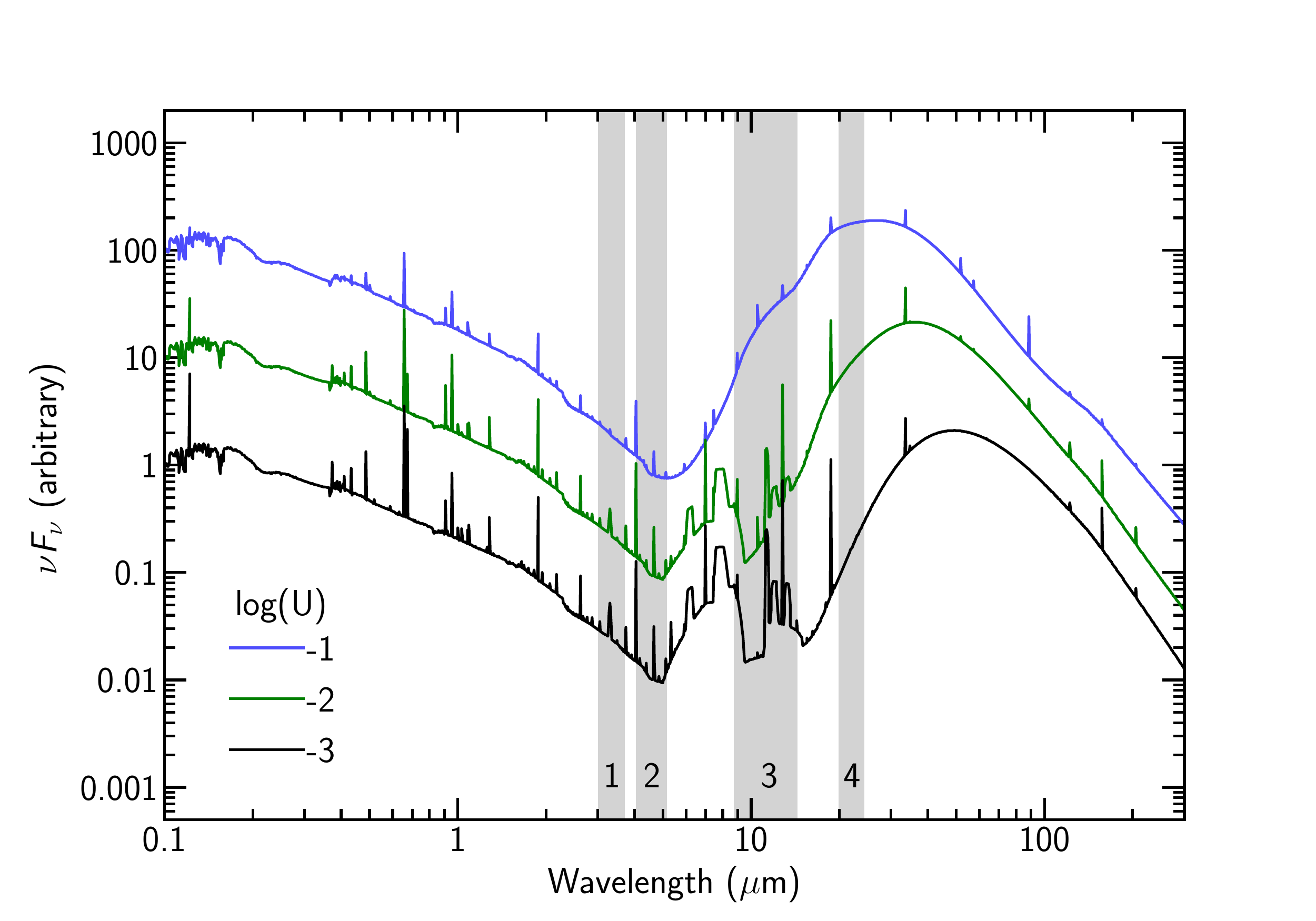}}
\caption{The transmitted continuum at a stopping column density of  $\log(N_\mathrm{H}/\textrm{cm}^{-2})=21$ for the pure starburst for the solar metallicity and standard gas density of $n_\mathrm{H}=300$~cm$^{-3}$ simulation for various ionization parameters, as indicated in the legend . The shaded region denotes the wavelength range of the first four WISE bands. Note that as $U$ increases, the dust temperature increases, and the strength of the PAH features decreases. At very high ionization parameters, the silicate feature appears in emission. These effects all contribute to changes in the resulting mid-infrared color. The three SEDs were offset by a factor of 10 from each other for clarity. Note that the apparent saturation in the figure is a plotting artifact related to the resolution of the data output.}
\label{transmitted_SED_Uvary}
\end{figure}

The shape of the infrared SED also depends strongly on the ionization parameter, with higher ionization parameters resulting in hotter dust and a corresponding shift in the infrared peak in the SED to shorter wavelengths as $U$ increases as can be seen in Figure ~\ref{transmitted_SED_Uvary}. This is because changes in $U$ result from changing the inner radius of the cloud, which affects the dust temperature.  Note also that the PAH and silicate features change significantly with ionization parameter. At high ionization parameters, a greater fraction of the gas is ionized for a given column density.  Since the PAHs are modeled to be destroyed within the ionized region, higher ionization parameters result in less prominent PAH features for a given column density.  Also, because a greater fraction of the gas is ionized with increasing U, the silicate feature appears in emission for the highest ionization parameters.  

\par
\begin{figure}
\noindent{\includegraphics[width=8.7cm]{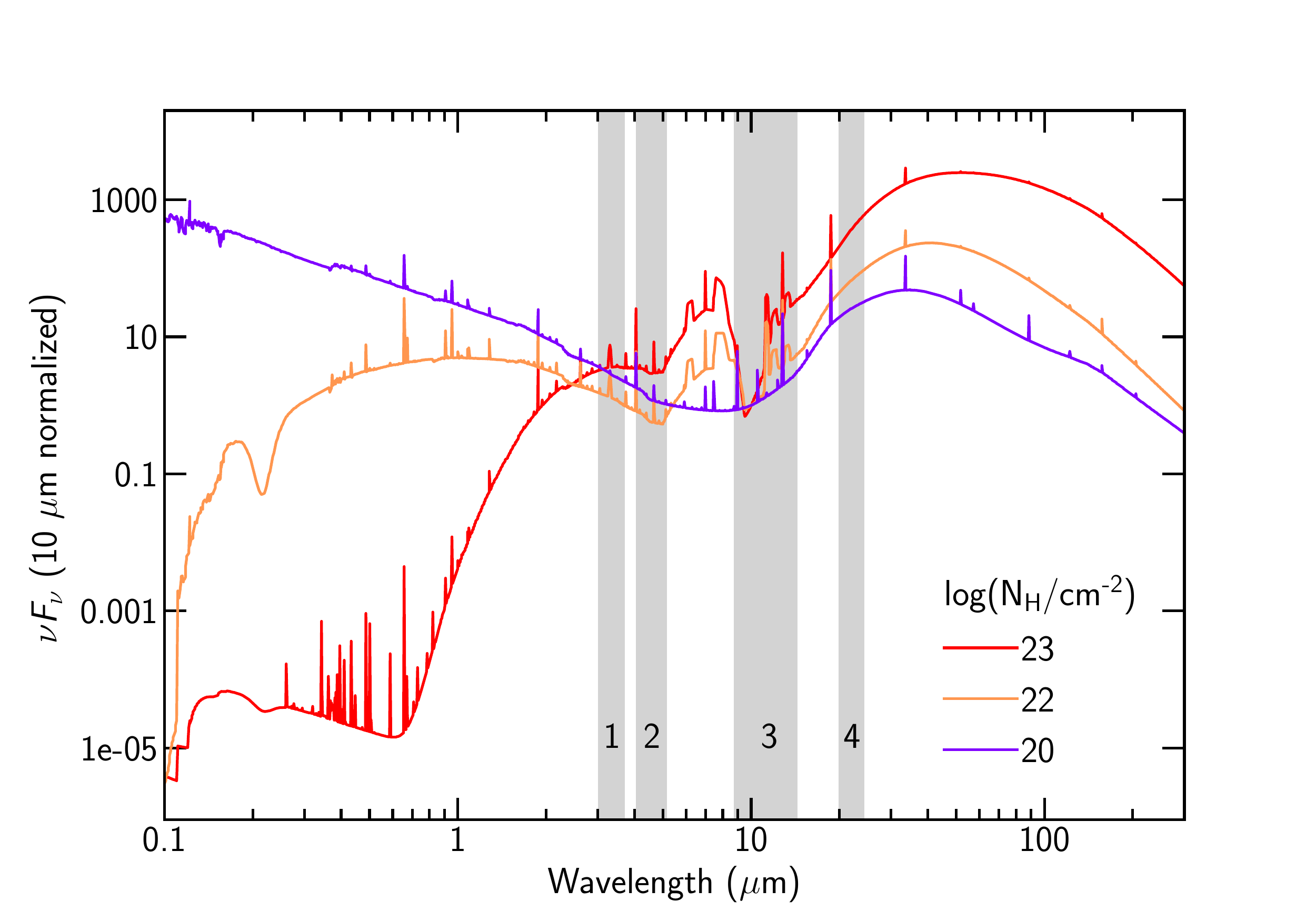}}
\caption{The transmitted continuum for the pure starburst model for the solar metallicity and standard gas density of $n_\mathrm{H}=300$~cm$^{-3}$ case for a selection of column densities indicated in the legend. The shaded region denotes the wavelength range of the first four WISE bands. Higher column densities result in cooler dust, and the transmitted infrared SED shifts to longer wavelengths, and more prominent PAH emission. The depth of the silicate feature also increases with $N_\mathrm{H}$.}
\label{transmitted_SED_NHvary}
\end{figure}

 
The shape of the transmitted SED is also strongly dependent on the stopping criterion adopted in the simulation. As $N_\mathrm{H}$ increases, the flux at short wavelengths gets attenuated and the peak of the infrared SED shifts to longer wavelengths, corresponding to cooler dust, as can be seen in Figure~\ref{transmitted_SED_NHvary}. Moreover, since PAHs are destroyed within the HII region and increase in abundance within the molecular region, the PAH features become more prominent with increasing $N_\mathrm{H}$, as the calculation extends further into the PDR. In addition, the depth of the 9.7~\micron silicate feature increases with increasing column density.  
\par
We emphasize that the strength of the PAH and silicate features depend in complex ways on the ionizing SED, the ionization parameter, and the stopping column density. Since these features fall within the WISE bands, their strength strongly affect the mid-infrared colors.It is therefore simplistic to assume that the infrared colors are only affected by the dust temperature, as was recently done by \citet{hainline2016}.

\par

\subsection{ The Dependence of Mid-Infrared Colors on Model Parameters}

\begin{figure}
\noindent{\includegraphics[width=8.7cm]{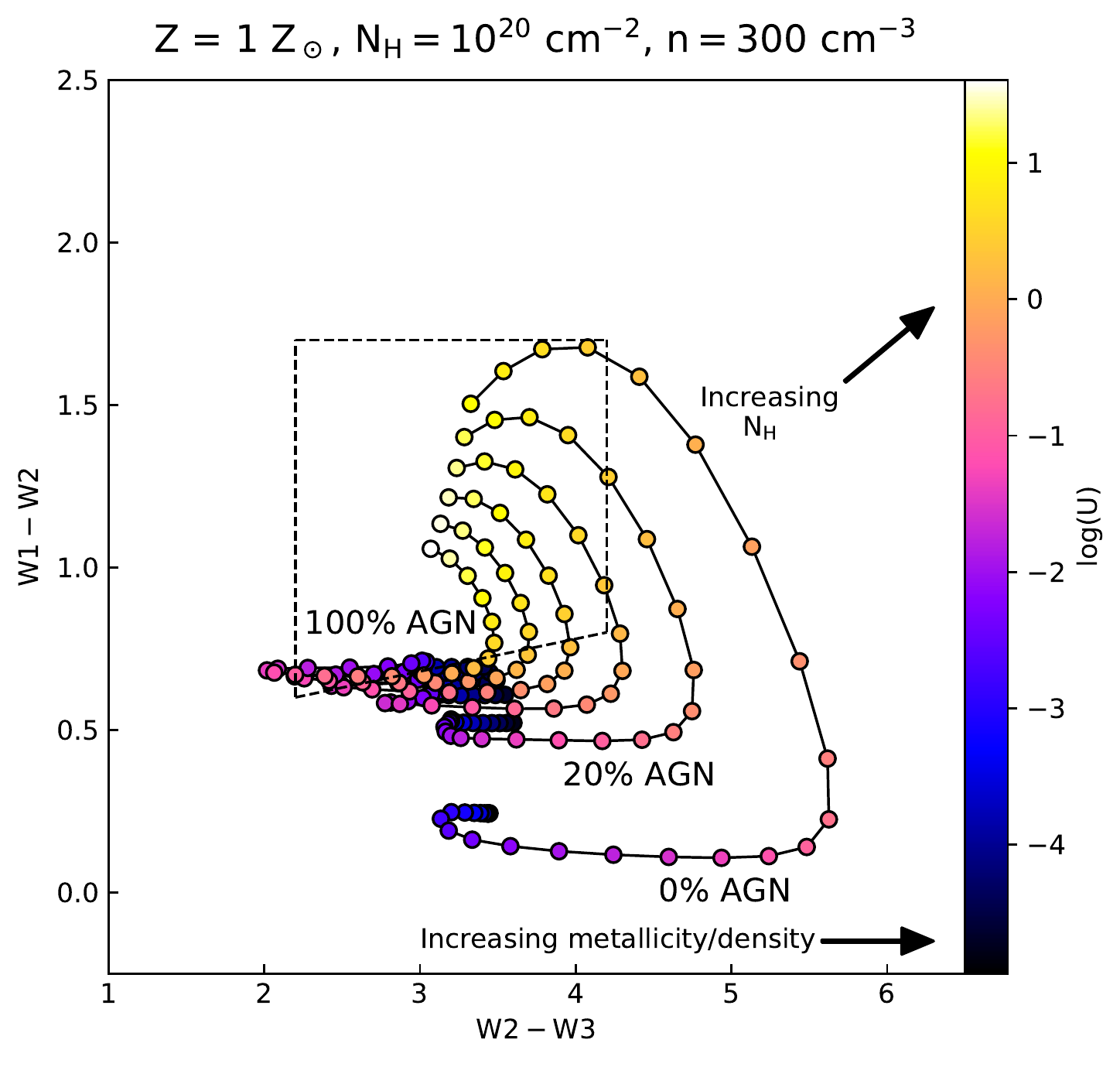}}
\caption{WISE color-color diagram for the solar metallicity models for the standard gas density $n_\mathrm{H}=300$~cm$^{-3}$ and a column density of $\log(N_\mathrm{H}/\textrm{cm}^{-2})=20$ with varying contributions of the AGN SED. Each track plotted shows different AGN contributions, ranging from 0\% (lowest track) to 100\% (highest track). The ionization parameter is shown by the auxiliary axis on the right side of the plot. The dependence of the colors on increasing column density, gas density, and metallicity are indicated by the directions of the arrows.The dashed black line shows the widely used AGN demarcation region from \citet{jarrett2011}.}
\label{solarwisecolor}
\end{figure}

The changes in the transmitted continuum with model parameters strongly affect the mid-infrared colors.  In Figure \ref{solarwisecolor}, we plot the resulting WISE colors on a color-color diagram for the solar metallicity models.  The ionization parameter is displayed by the auxiliary axis color bar. The AGN demarcation region from \citet{jarrett2011} is shown by the dotted black square in the figure.   As can be seen, the mid-infrared colors are strongly dependent on the contribution of the AGN to the incident radiation field and the ionization parameter, with the pure starburst model falling well below the AGN demarcation region for all but the highest ionization parameters.  In contrast, the mid-infrared colors for the models in which the AGN contributions dominate fall within the AGN region for a large range of ionization parameters, as expected.  Note that for the pure starburst model,  the $W1-W2$ color increases with ionization parameter and can exceed 0.8, the color cut employed by \citet{stern2012}, at moderate ionization parameters. However, the $W2-W3$ color for these models are significantly redder than the AGN box displayed in the figure.Note that for the most extreme ionization parameters, the colors get bluer for a given column density.This is likely the effect of dust absorption of the UV photons in the HII region.  At extreme ionization parameters, the dust absorbs more of the photons with energies $>$ 13.6~eV than does the gas.  As a result, the shape of the transmitted continuum changes for a given column density, which in turn affects the mid-infrared colors. 

\par
The stopping criterion also has a significant effect on the mid-infrared colors, with the colors shifting toward the upper right of the plot with increasing $N_\mathrm{H}$, as indicated by the direction of the arrow in Figure \ref{solarwisecolor}. For the AGN models, the mid-infrared colors stay within the AGN box displayed in the figure for a large range of column densities, in contrast to the pure starburst models. In addition to $N_\mathrm{H}$, the metallicity plays a strong role on the mid-infrared colors. While the SED for the low metallicity starburst is harder, the total dust abundance is reduced.  The net result is that higher column densities are required to obtain the same WISE colors at a given ionization parameter for the pure starburst model compared with the solar metallicity models. For a given ionization parameter and column density, the mid-infrared colors shift to the right with increasing metallicity, as indicated by the direction of the arrow in Figure \ref{solarwisecolor}.  Finally, the gas density also affects the shape of the mid-infrared continuum.  This is because a higher density model implies a higher photon flux for the same ionization parameter, allowing for lower ionization parameters to generate WISE colors consistent with the lower density and higher ionization parameter models.
\par
  
\subsection{ Extreme Starbursts as AGN Imposters?}

Our simulations demonstrate that a pure starburst model can easily produce $W1-W2$ colors similar to those of AGNs, for a significant range of parameter space. Our simulations also show that in principle, an extreme starburst can also mimic an AGN in its mid-infrared colors based on three band color cuts used in the literature for a narrow range of parameter space as shown in Figure \ref{solarwisecolor}.  Our integrated modeling approach allows us to predict both the line and emergent continuum.  Any model that replicates the observed mid-infrared continuum, must also be able to reproduce the observed optical emission line ratios of star forming galaxies and AGNs.  In this section, we discuss the region of parameter space in which the mid-infrared colors are consistent with commonly used AGN mid-infrared color-cuts and simultaneously consistent with the optical emission line ratios of star forming galaxies.
\par
\begin{figure*}[]

\centering

\begin{tabular}{cc}

\includegraphics[width=0.42\textwidth]{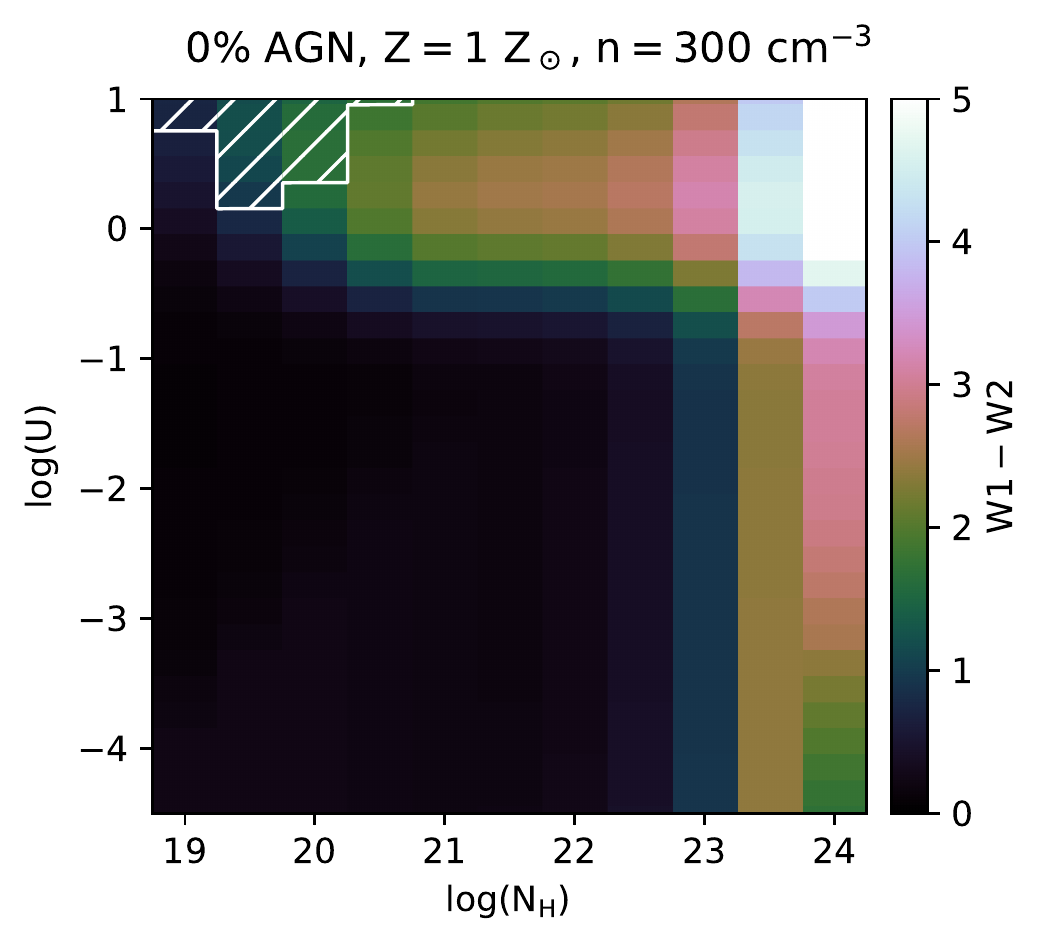} & \includegraphics[width=0.42\textwidth]{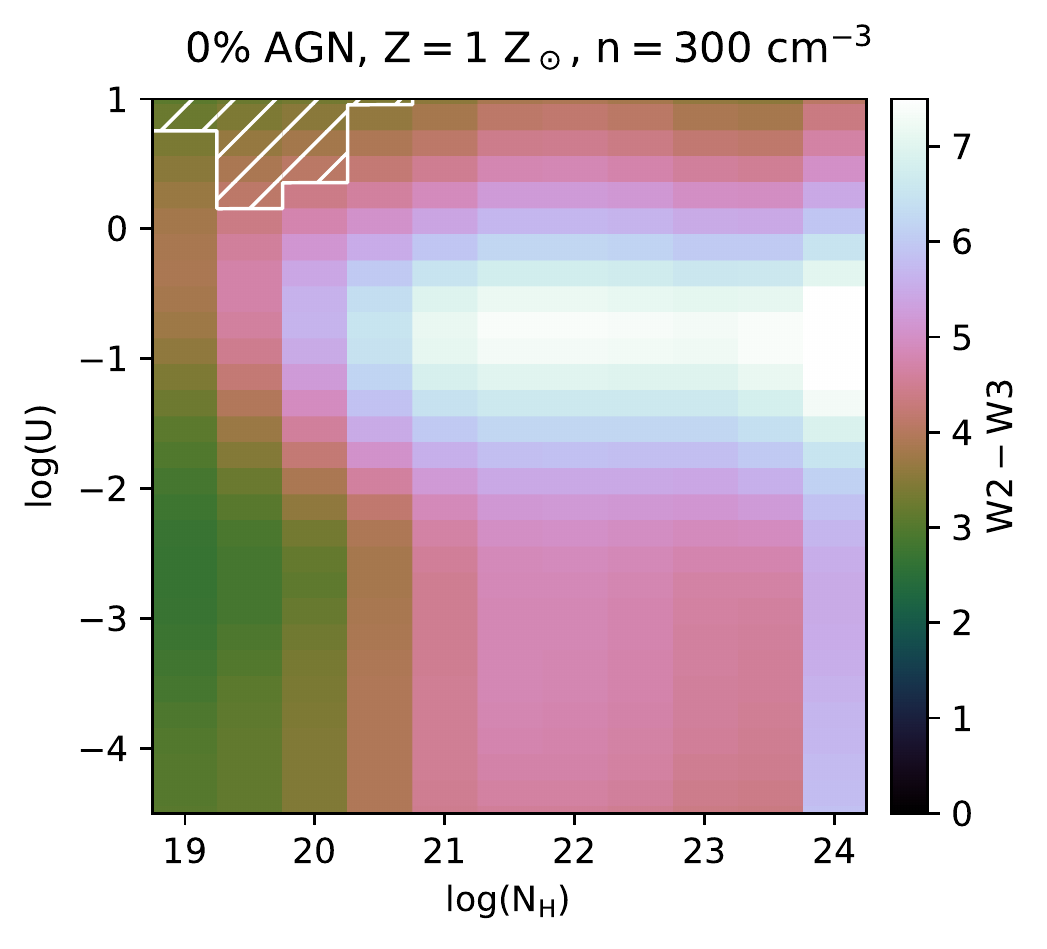} \\

\includegraphics[width=0.42\textwidth]{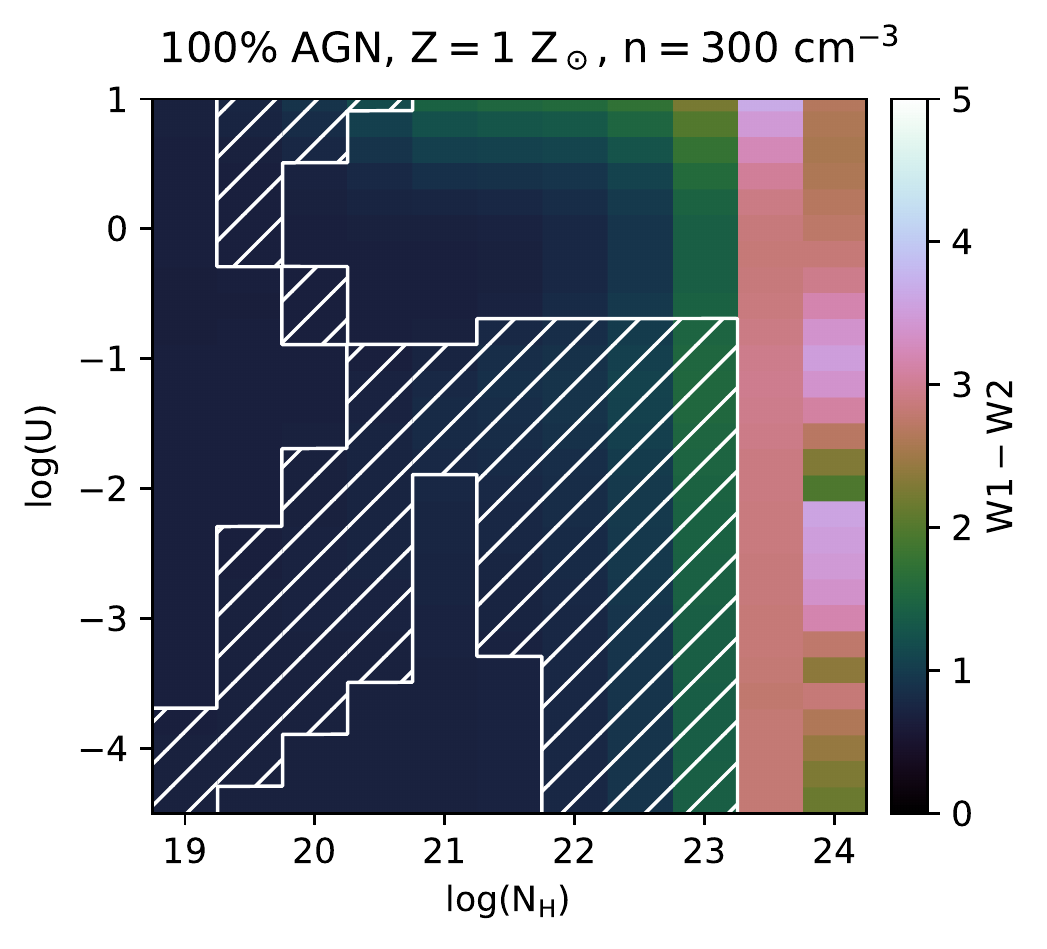} & \includegraphics[width=0.42\textwidth]{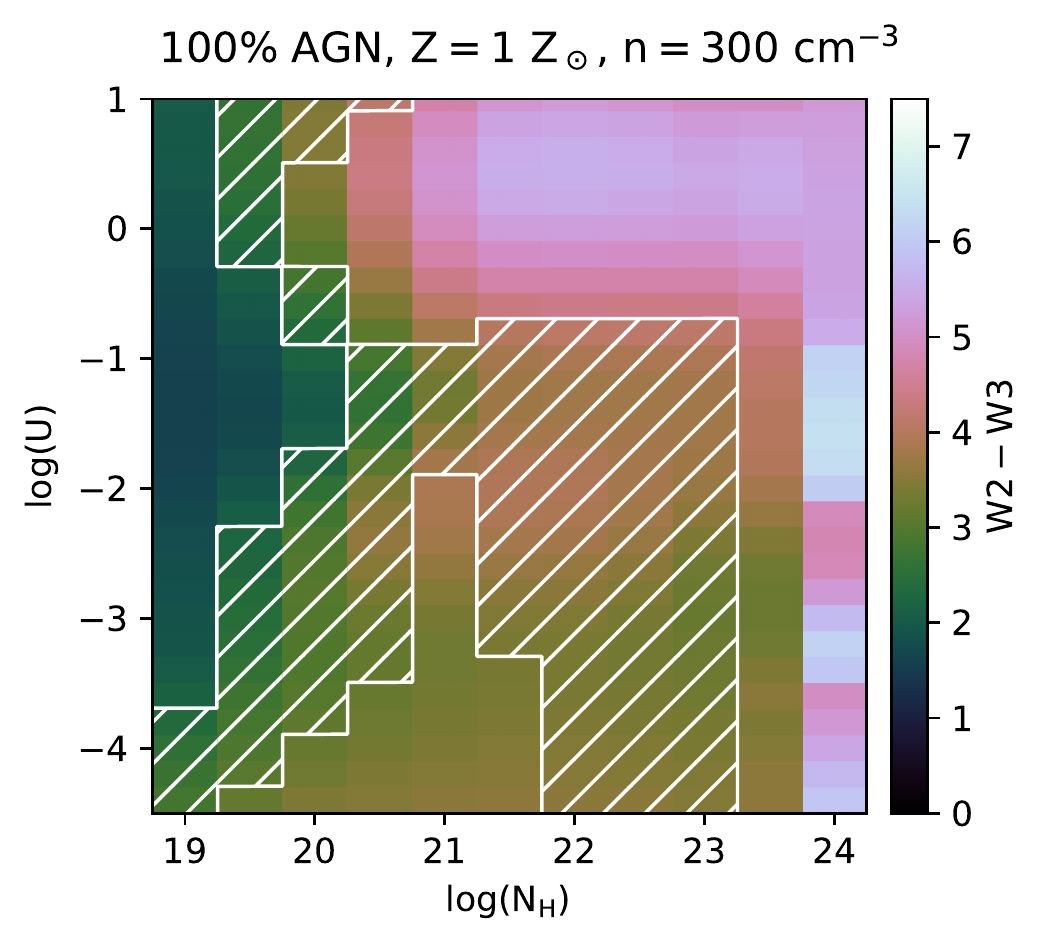} \\

\end{tabular}
\caption{The predicted WISE colors for the solar metallicity pure starburst and 100\% AGN models for the standard gas density $n_\mathrm{H}=300$~cm$^{-3}$ model as a function of ionization parameter and stopping column density ($\log(N_\mathrm{H}\textrm({cm}^{-2}$)). The shaded region denotes the region in which the colors fall within the AGN demarcation region from \citet{jarrett2011} . Note that the region of parameter space in which the AGN colors fall within the \citep{jarrett2011} box is much larger than is seen for the pure starburst model, as expected.}
\label{wisecolor_parameters}
\end{figure*}
\begin{figure*}[]

\centering

\begin{tabular}{cc}

\includegraphics[width=0.42\textwidth]{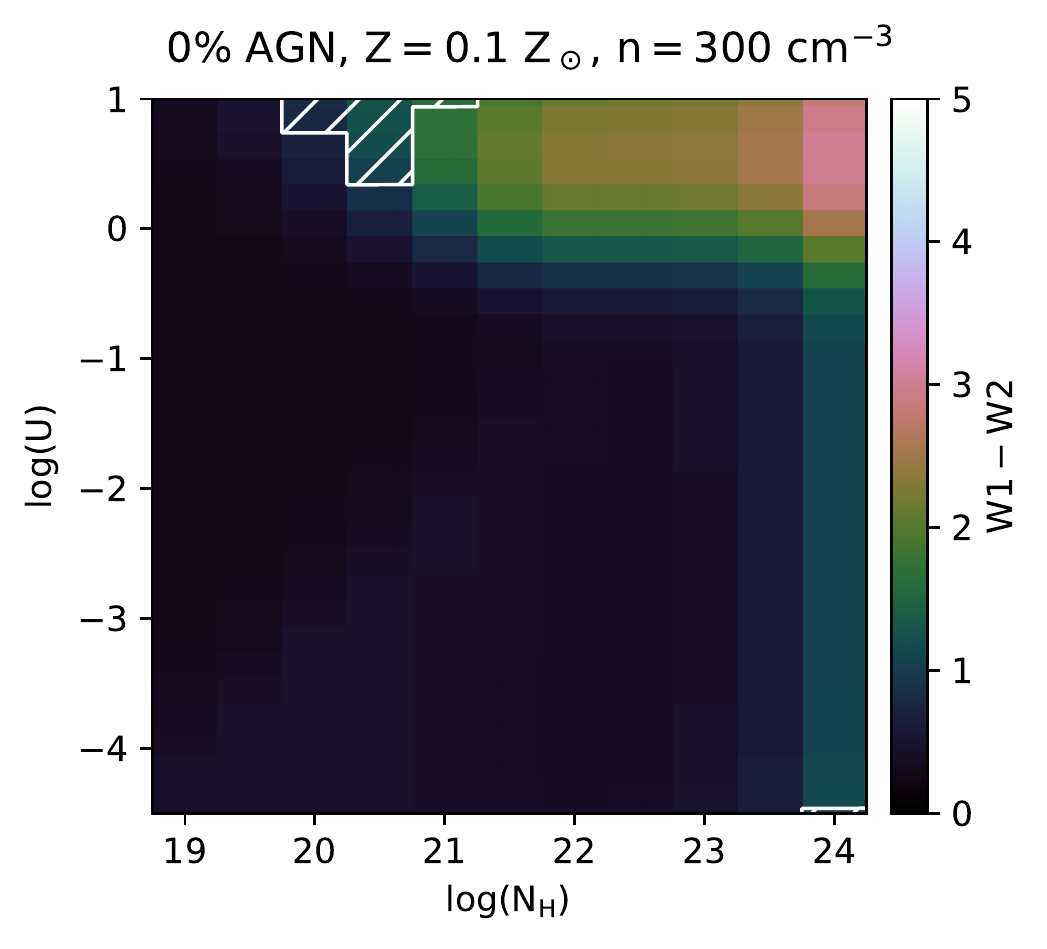} & \includegraphics[width=0.42\textwidth]{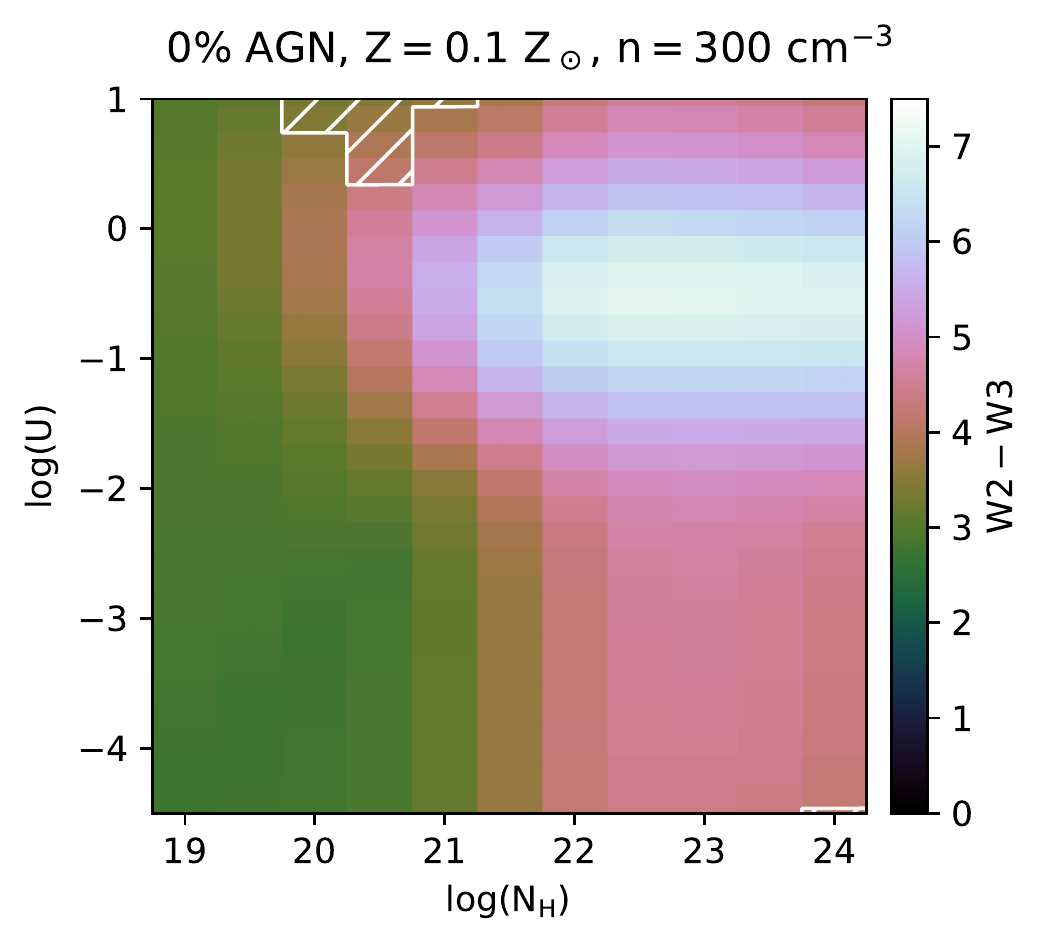} \\

\includegraphics[width=0.42\textwidth]{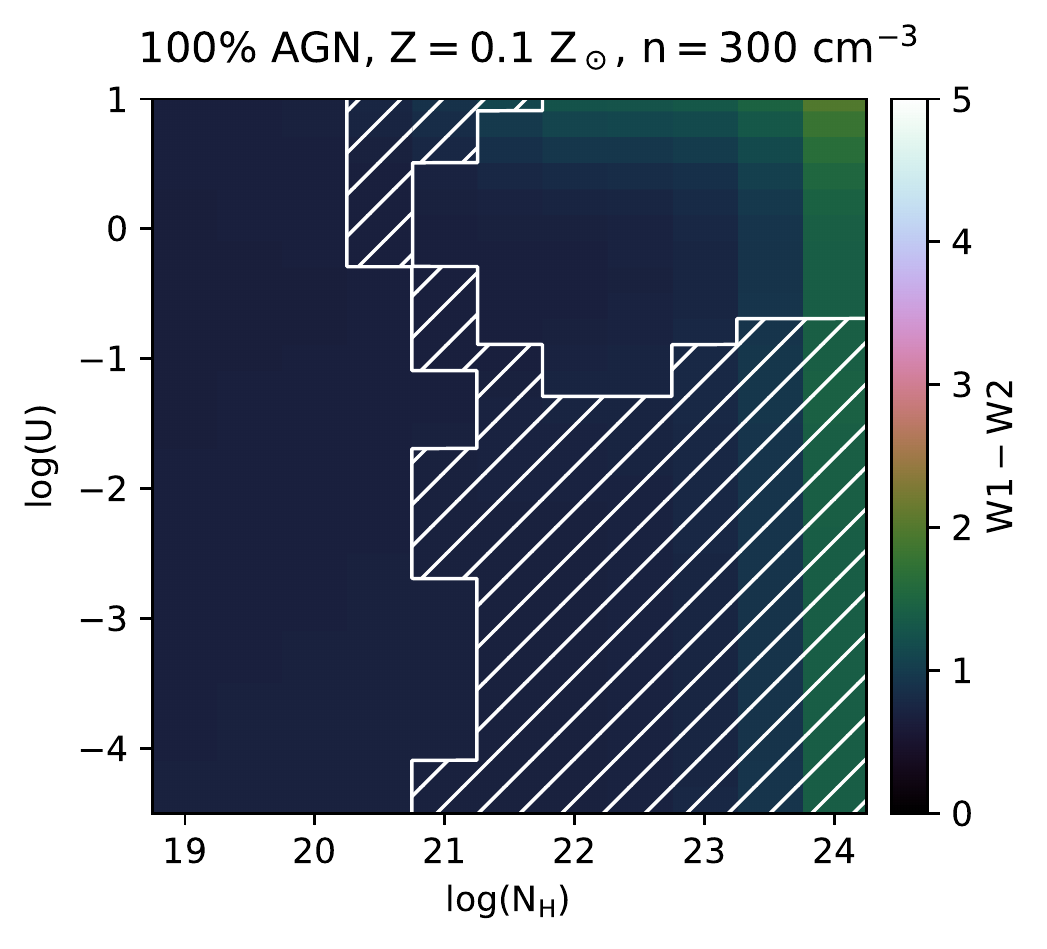} & \includegraphics[width=0.42\textwidth]{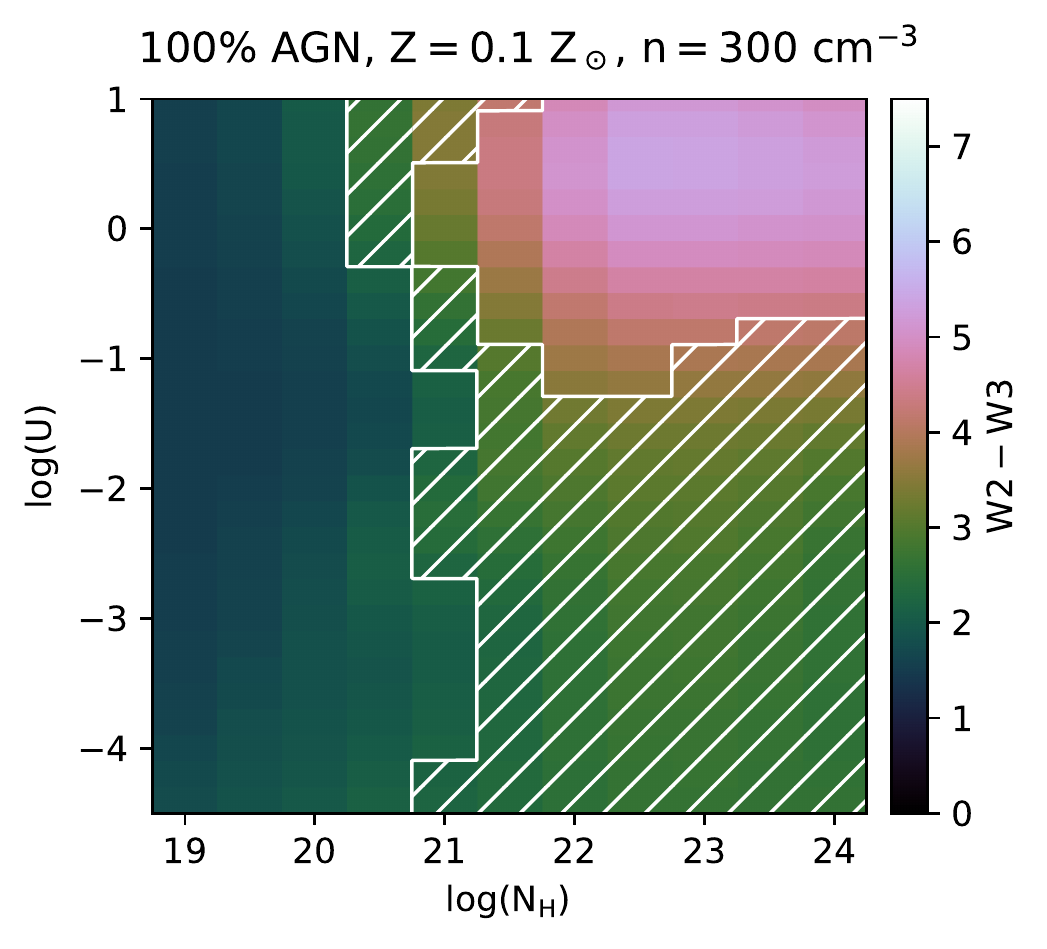} \\

\end{tabular}
\caption{The predicted WISE colors for the low metallicity pure starburst and 100\% AGN models for the standard gas density $n_\mathrm{H}=300$~cm$^{-3}$ model as a function of ionization parameter and stopping column density($\log(N_\mathrm{H}\textrm({cm}^{-2}$)). The shaded region denotes the region in which the colors fall within the AGN demarcation region from \citet{jarrett2011} . Note that the region of parameter space in which the AGN colors fall within the \citep{jarrett2011} box is much larger than is seen for the pure starburst model, as expected. For the low metallicity models, the shaded region moves slightly to the right, since higher column densities are required to replicate the same WISE colors of the solar metallicity models.}
\label{wisecolor_parameters2}
\end{figure*}

In Figure \ref{wisecolor_parameters}, we show the mid-infrared colors of the pure starburst and 100\% AGN models for the solar metallicity and standard gas density model as a function of ionization parameter and stopping column density.  The shaded gray region denotes the region of parameter space in which the colors fall within the AGN demarcation region from \citet{jarrett2011}. As can be seen, it is possible for a pure starburst to produce redder 
$W1-W2$ WISE colors  ($W1-W2 > 0.8$) than the observed colors of the fast majority of star forming galaxies (e.g., see Figure 12 in \citep{yan2013}) for a significant range of parameter space with moderately high ionization parameters or high column densities. Thus single color cuts such as the commonly employed $W1-W2 > 0.8$ color cut from \citet{stern2012} can in principle be created by pure starbursts. However, these models in general produce redder $W2-W3$ colors. The range of parameter space for the pure starburst models where both the $W1-W2$ and $W2-W3$ colors fall within the \citet{jarrett2011} AGN demarcation region is narrow.  There is a region in the upper left portion of the parameter space (low column density, high ionization parameter) that reproduces both regions, but the ionization parameters are extreme and inconsistent with the optical line ratios of any observed star forming galaxy, which typically are between $-3.2<\log(U)<-2.9$ for local HII regions \citep{dopita2000} and local star forming galaxies \citep{moustakas2010}. Although some high redshift galaxies may have higher ionization parameters \citep{pettini2001,maiolino2008}, the most anomalously high ionization parameter in a low metallicity dwarf at $z=2.3$ is still only $\log(U)=-1$ \citep{erb2010}, well below the ionization parameters required to produce mid-infrared colors in the \citet{jarrett2011} box. In contrast, the AGN model can reproduce both sets of WISE colors for a much wider range of parameter space. In Figure \ref{wisecolor_parameters2}, we show the mid-infrared colors of the pure starburst and 100\% AGN models this time for the low metallicity and standard gas density model as a function of ionization parameter and stopping column density. Again, while the pure starburst can mimic an AGN based on the \citet{jarrett2011} AGN color cut, this is for a very narrow range of parameters corresponding to extreme ionization parameters that are inconsistent with the observed optical emission line ratios of star forming galaxies. Note that at lower metallicity, the decrease in the dust abundance results in higher column densities for the same temperature dust. Thus the shaded region in Figure \ref{wisecolor_parameters} moves slightly to the right compared with Figure \ref{wisecolor_parameters2}.

\par
\begin{figure*}[]

\centering

\begin{tabular}{cc}

\includegraphics[width=0.42\textwidth]{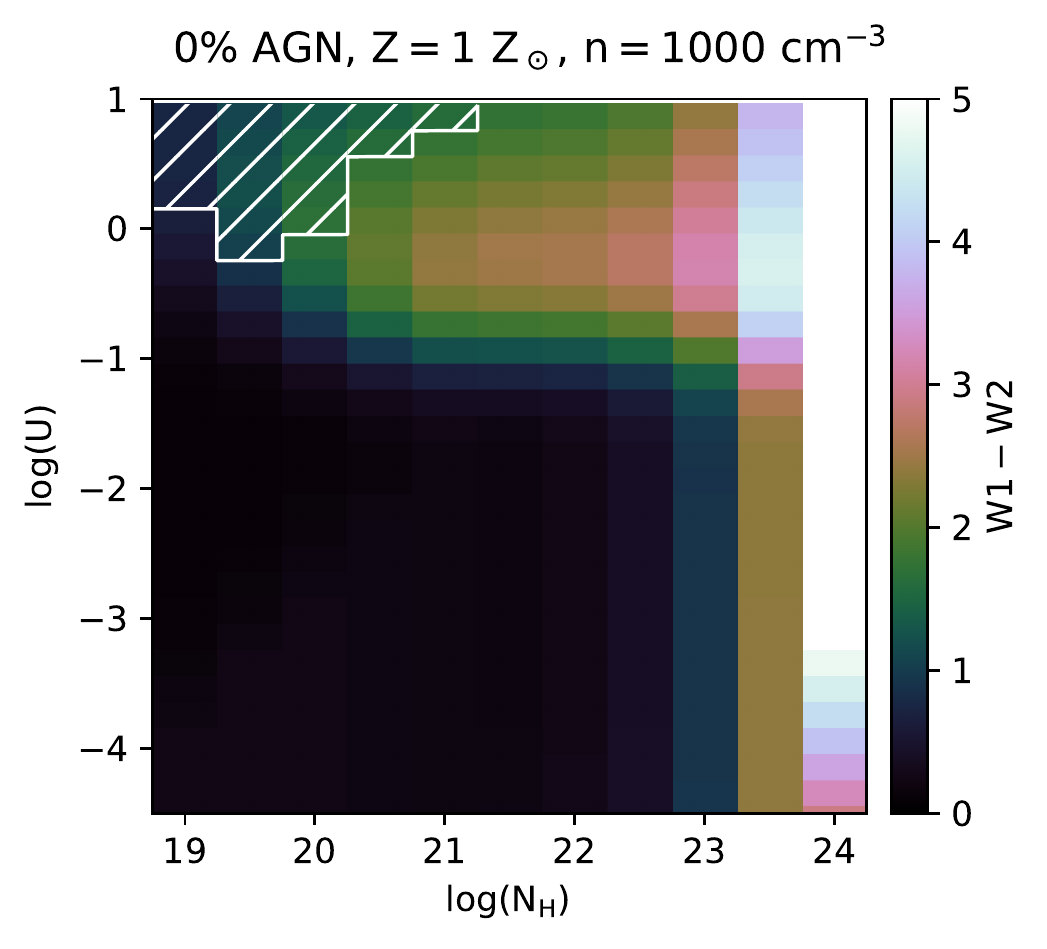} & \includegraphics[width=0.42\textwidth]{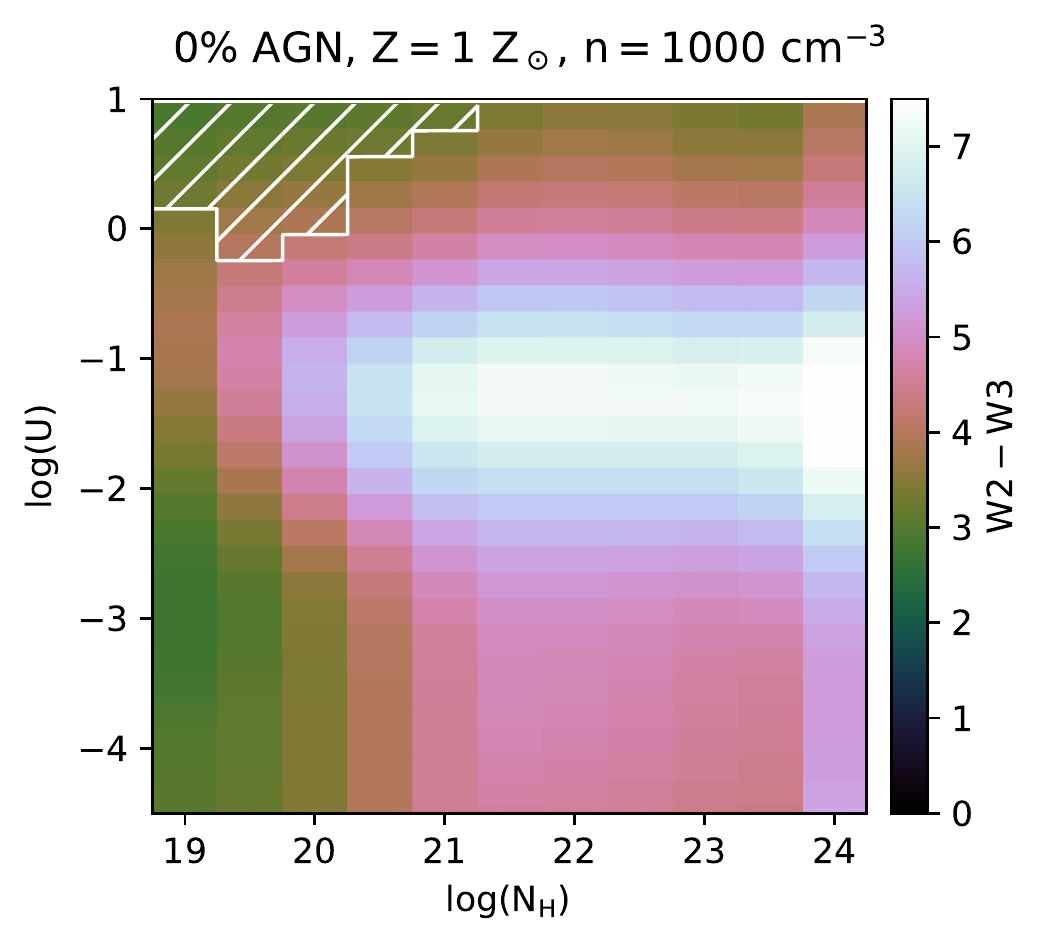} \\

\includegraphics[width=0.42\textwidth]{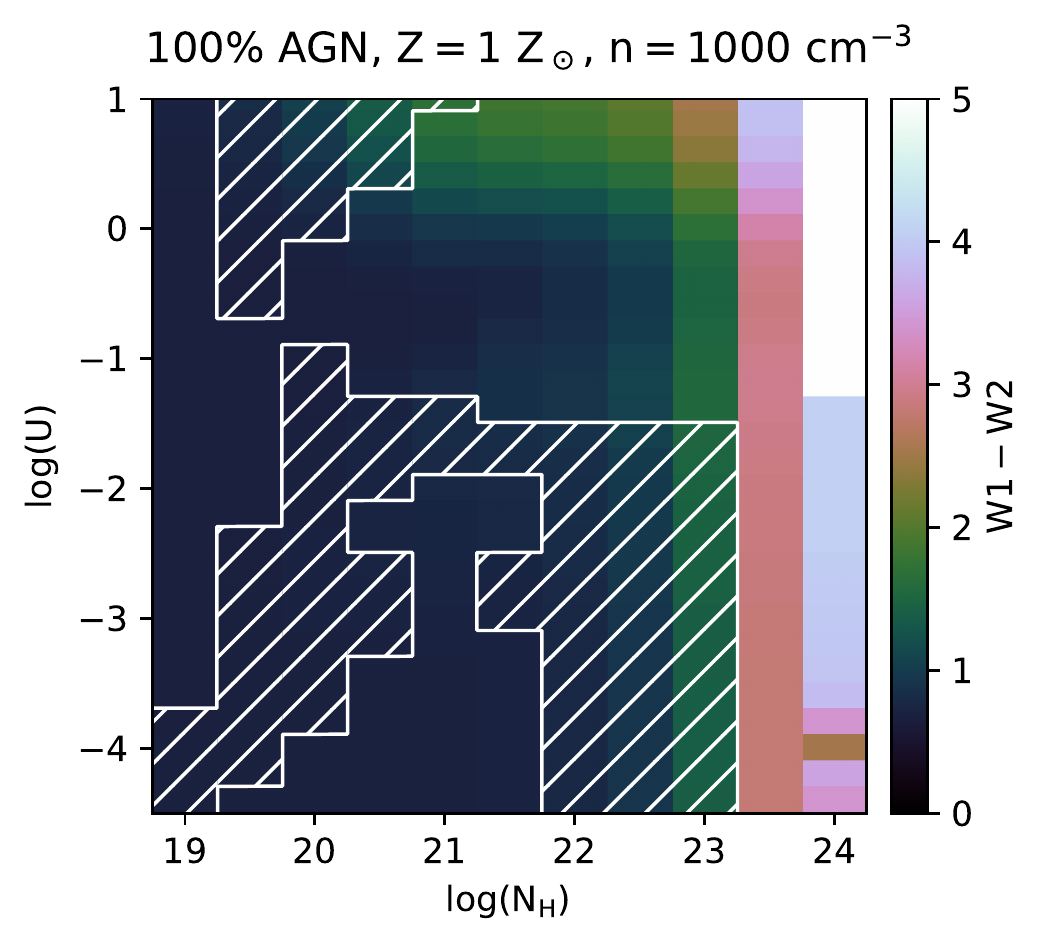} & \includegraphics[width=0.42\textwidth]{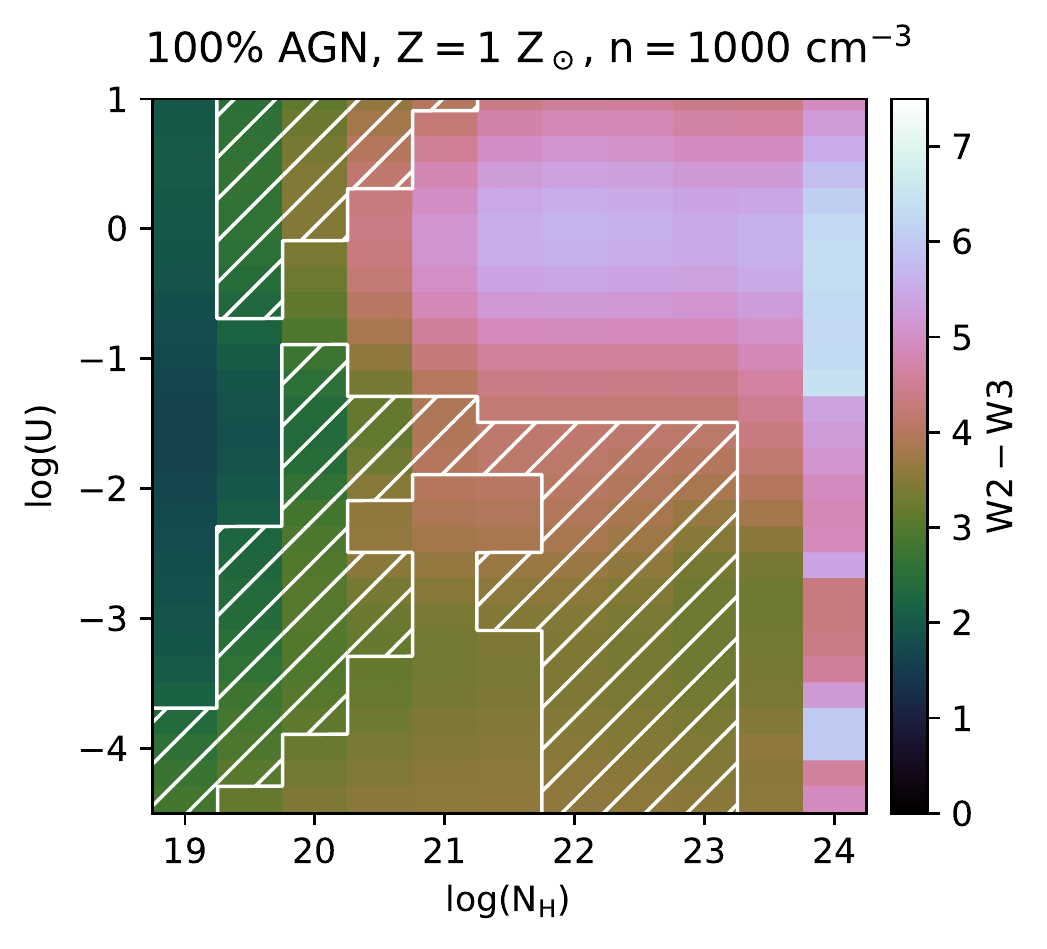} \\

\end{tabular}
\caption{The predicted WISE colors for the solar metallicity pure starburst and 100\% AGN models for the high gas density $n_\mathrm{H}=1000$~cm$^{-3}$ model as a function of ionization parameter and stopping column density ($\log(N_\mathrm{H}\textrm({cm}^{-2}$)). The shaded region denotes the region in which the colors fall within the AGN demarcation region from \citet{jarrett2011} . Note that the region of parameter space in which the AGN colors fall within the \citep{jarrett2011} box is much larger than is seen for the pure starburst model, as expected.}
\label{wisecolor_parameters3}
\end{figure*}
\begin{figure*}[]

\centering

\begin{tabular}{cc}

\includegraphics[width=0.42\textwidth]{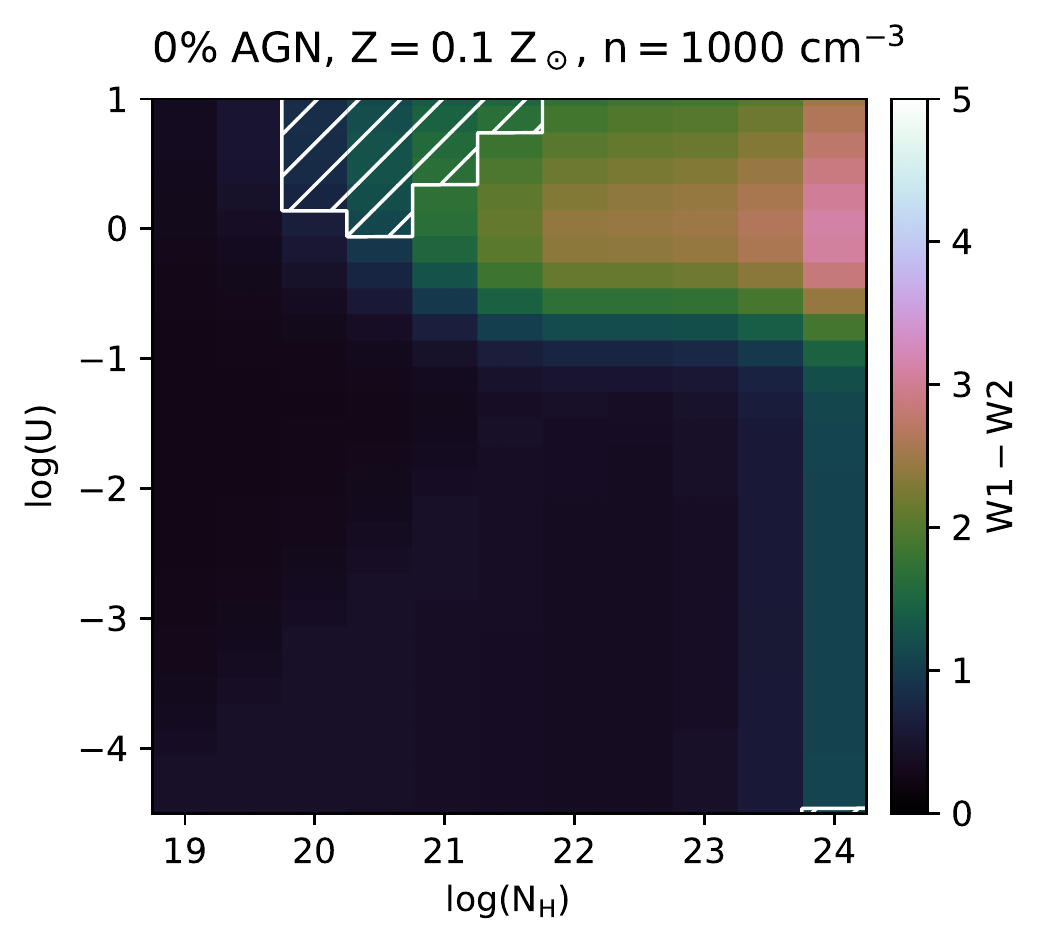} & \includegraphics[width=0.42\textwidth]{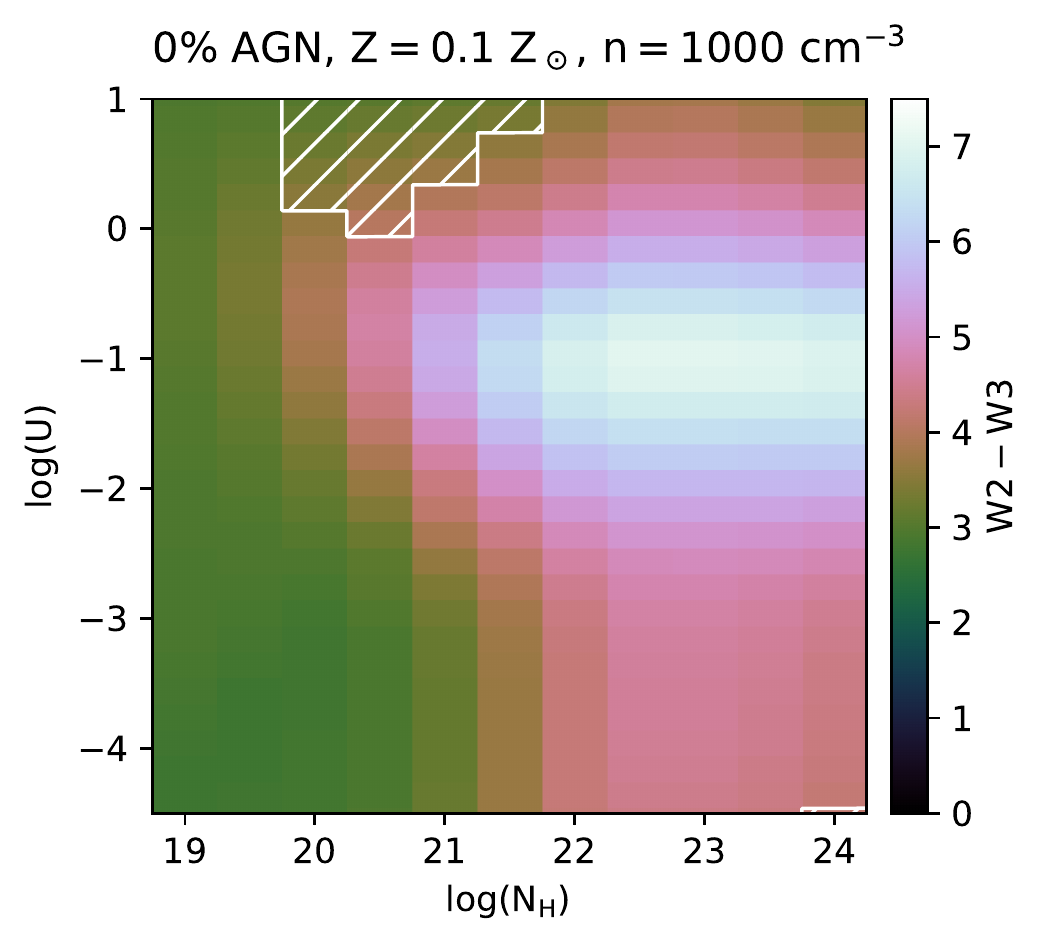} \\

\includegraphics[width=0.42\textwidth]{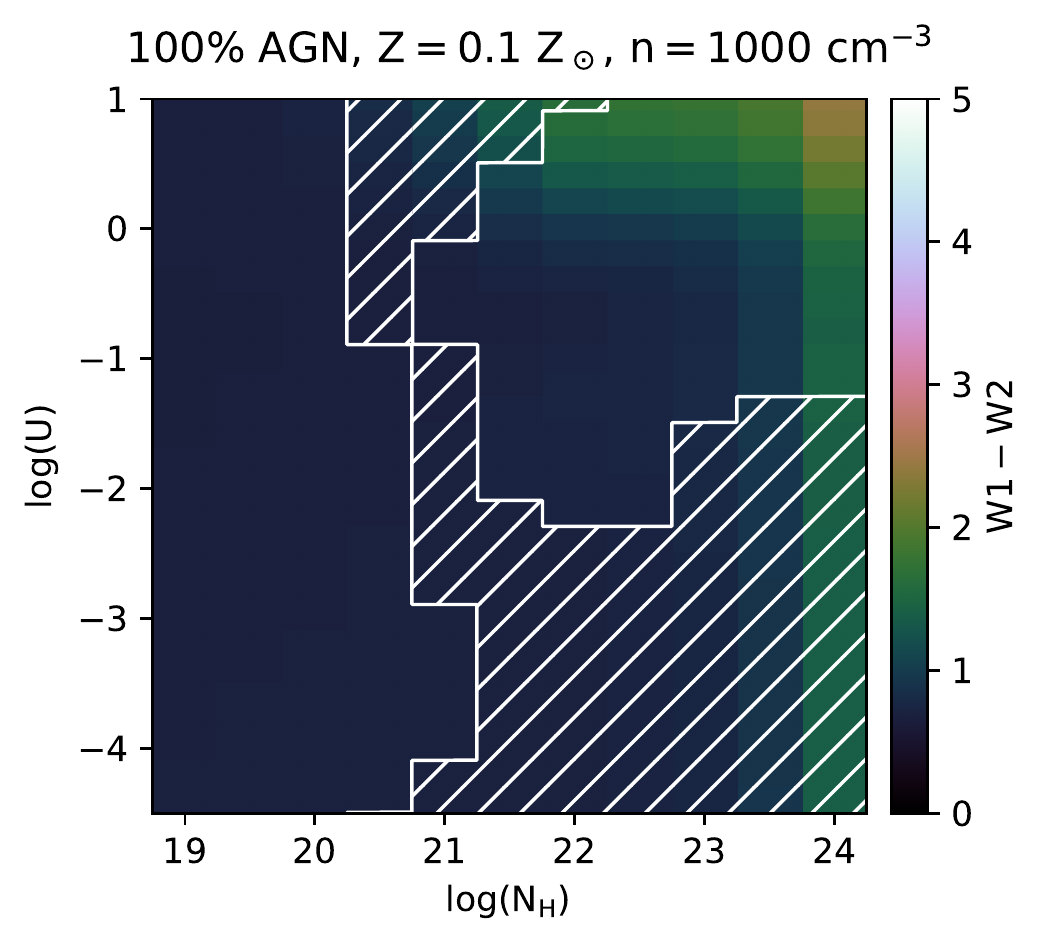} & \includegraphics[width=0.42\textwidth]{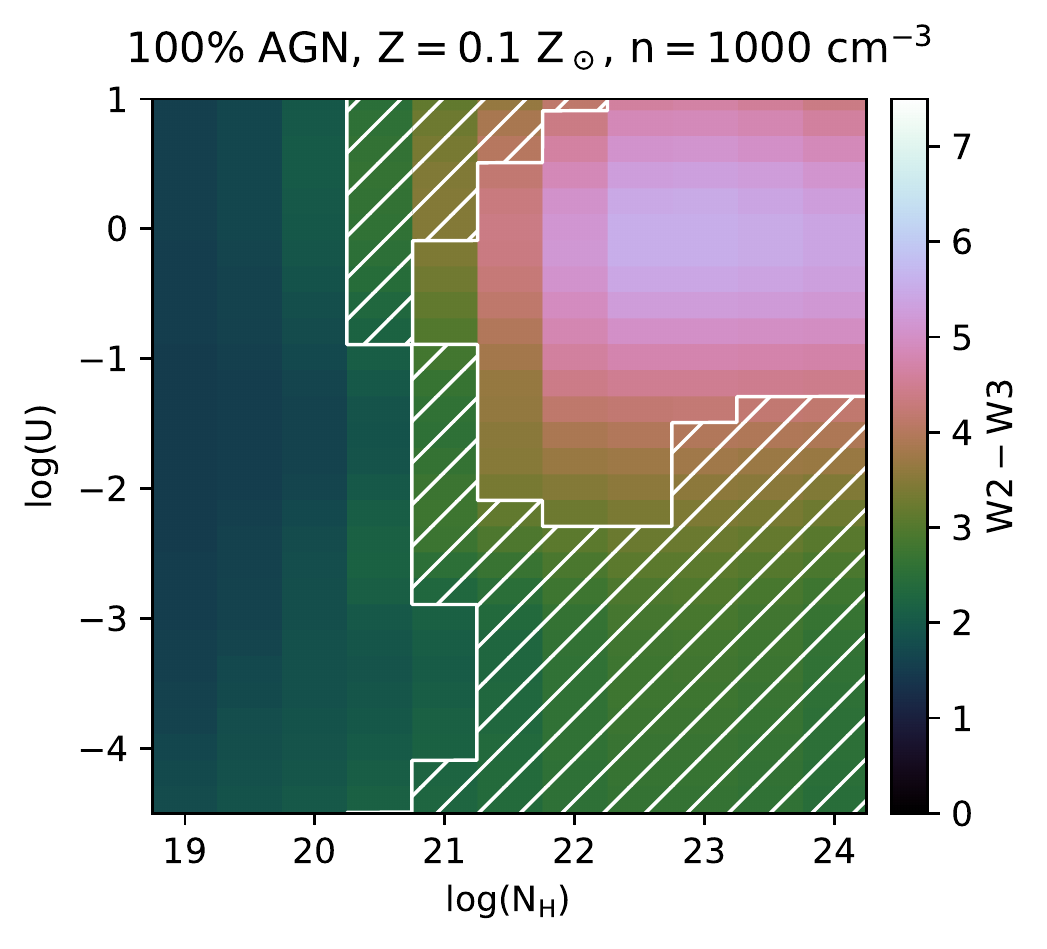} \\

\end{tabular}
\caption{The predicted WISE colors for the low metallicity pure starburst and 100\% AGN models for the high gas density $n_\mathrm{H}=1000$~cm$^{-3}$ model as a function of ionization parameter and stopping column density($\log(N_\mathrm{H}\textrm({cm}^{-2}$)). The shaded region denotes the region in which the colors fall within the AGN demarcation region from \citet{jarrett2011} . Note that the region of parameter space in which the AGN colors fall within the \citep{jarrett2011} box is much larger than is seen for the pure starburst model, as expected. For the low metallicity models, the shaded region moves slightly to the right, since higher column densities are required to replicate the same WISE colors of the solar metallicity models.}
\label{wisecolor_parameters4}
\end{figure*}

 In Figures  \ref{wisecolor_parameters3} and  \ref{wisecolor_parameters4}, we show the effects of increasing gas density on the mid-infrared colors. The overall effect of higher density is to lower the ionization parameter needed to produce the same infrared SED.  This is because increasing the density by a factor of 30 means that the ionization parameter needed to produce the same ionizing flux is lowered by a factor of 30.  Therefore, the dust temperature can be higher for the high density models for lower ionization parameters. Thus, the infrared SED for an ionization parameter of $\log(U)=-1$ and $n_\mathrm{H}=300$~cm$^{-3}$ will be similar to the infrared SED for an ionization parameter of $\log(U)=-2.5$ and $n_\mathrm{H}=10^{3}$ cm$^{-3}$.  Thus  the region of parameter space in which the mid-infrared colors are in consistent with the \citet{jarrett2011} AGN demarcation for the high density pure starburst model extends to lower ionization parameter compared to the lower gas density models. However, the required ionization parameters are still in excess of what is typically observed. A combination of extreme ionization parameters and gas densities would be required. Note that only in extremely active star forming galaxies and ULIRGs, the gas densities can get as high as  $n_\mathrm{H}=1000$~cm$^{-3}$  \citep{kewley2001b,armus2004}.

\par

\par

\section{Discussion}

\subsection{ A Theoretical AGN Mid-infrared Color-cut}

In contrast to mid-infrared color-cuts that are based on empirical templates, our integrated modeling approach allows us to understand how changing various physical parameters affects the shape of the infrared continuum, the PAH and silicate features, and hence the mid-infrared colors. Moreover, these models not only demonstrate how the position on mid-infrared color-color plots trace the input radiation field and ISM conditions, but they simultaneously predict the emission line spectrum of the gas, and allow for both the mid-infrared continuum and various emission line ratios to be self-consistently modeled.
\par
In this work, our integrated modeling effort has demonstrated from purely theoretical grounds that an extreme starburst can mimic an AGN in its mid-infrared colors. However, the range of parameter space in which the stringent three-band color cut from \citep{jarrett2011} is reproduced is narrow, and corresponds to extremely high ionization parameters or gas densities that are inconsistent with the observed optical emission line ratios of currently observed star forming galaxies. Mid-infrared color cuts based on empirical galaxy templates and observed colors of AGNs identified through previous selection methods can potentially either have star forming galaxies scatter into the defined ``AGN region'', or be too restrictive and unnecessarily exclude less luminous or highly obscured AGNs, since a galaxy may have an obscured and/or a lower luminosity AGN that is not identified either through optical or X-ray selection. Based on our theoretical calculations, we can construct a theoretical mid-infrared color cut that will exclude even the most extreme starburst that we have modeled in this work. For low redshift galaxies ($z < 0.2$), we define the following color-cut that can be adopted that will be free of contamination from any purely star-forming galaxy:

\begin{equation}
W_{12} \geq 0.52 \text{ and } W_{12} \geq 5.78W_{23}-24.50   
\end{equation}

\begin{figure}
\noindent{\includegraphics[width=8.7cm]{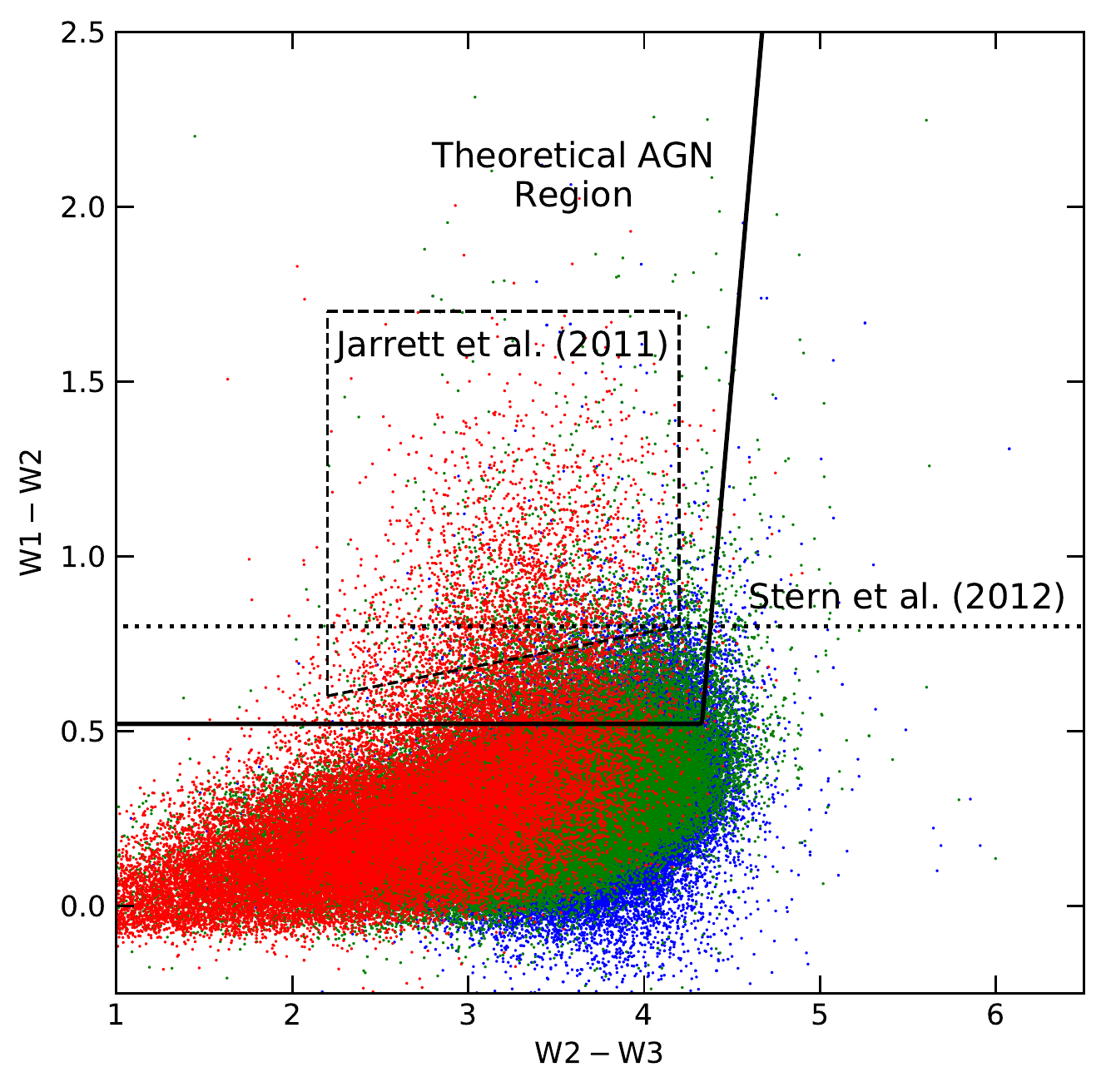}}
\caption{WISE color-color diagram showing the theoretical AGN demarcation region presented in this work, together with the \citet{jarrett2011} AGN box, and the $W1-W2 > 0.8$ color cut from \citet{stern2012}.  We also plot the colors of SDSS galaxies detected by WISE with signal to noise ratio greater than five in the first three WISE bands that also have H$\beta$, [OIII] $\lambda5007$, H$\alpha$, and [NII] $\lambda6584$ emission line fluxes detected with a signal to noise $> 3\sigma$ . Galaxies were identified as AGNs (red points) based on their optical line ratios following the classification scheme of \citet{kewley2001}. Galaxies were classified as star forming galaxies (blue points) based on the classification scheme from \citet{stasinska2006}, with composites representing those galaxies with emission line ratios between the two (green points). }
\label{agnregions}
\end{figure}

Our theoretical demarcation region includes approximately a factor of two times as many local galaxies ($z<0.2$) than does the \citet{jarrett2011} AGN box. Based on all galaxies with redshifts $z<0.2$ drawn from the Max Planck Institute for Astrophysics/Johns Hopkins University (MPA/JHU) collaboration,\footnote{\url{http://www.mpa-garching.mpg.de/SDSS/}} which provides a large database of galaxies with optical spectroscopic observations, we examined the optical emission line ratios of galaxies with detections in the WISE $W1$ and $W2$ bands with a signal-to-noise greater than $5\sigma$ and the $W3$ band with a signal-to-noise greater than $3\sigma$. Using only galaxies with H$\beta$, [OIII] $\lambda5007$, H$\alpha$, and [NII] $\lambda6584$ emission line fluxes detected with a signal to noise greater than $3\sigma$, 0.31\% of optically classified star forming galaxies following the \citet{stasinska2006} BPT demarcation also fall within the \citet{jarrett2011} AGN box, compared with 3.4\% that fall within our theoretical AGN region. Based on our modeling, these are likely to contain optically hidden AGNs and are prime targets for follow up multi-wavelength studies. Our theoretical AGN region is more inclusive and includes a factor of $\sim~2$ times as many optically identified AGNs in the MPA/JHU catalog than does the \citet{jarrett2011} AGN box. The theoretical mid-infrared color cut presented in this work, along with  \citet{jarrett2011} and \citet{stern2012} color cuts, and the infrared colors of SDSS galaxies are shown in Figure~\ref{agnregions}.
 
 \subsection{ Optically Classified Star-forming Galaxies with Mid-infrared Identified AGNs?}

There have been a number of recent studies that have revealed an unexpected population of optically classified star forming galaxies with extremely red mid-infrared colors suggestive of AGN activity  \citep[e.g.,][]{griffith2011,izotov2011,satyapal2014,sartori2015,secrestb2015,satyapal2016,oconnor2016, hainline2016} .   In many cases, these galaxies lack significant bulges or are low mass galaxies, a population that thus far  seemed to be quiescent based on optical studies.  Low mass and bulgeless galaxies have undergone a more secular evolution than bulge-dominated massive galaxies and therefore the mass distribution and occupation fraction of SMBHs in these galaxies contains clues about the original seed population, allowing us to discriminate between lower mass seeds formed from stellar remnants or massive seeds formed directly out of dense gas \citep[e.g.][]{vanwassenhove2010}. The study of black holes at the low bulge mass regime is therefore crucial to our understanding of both the origin of SMBHs and their growth and connection to galaxy evolution. 
\par
While the red mid-IR colors are highly suggestive of accretion activity in these optically quiescent low mass and bulgeless galaxies, and follow-up multiwavelength studies provide some compelling support for this scenario in a few cases \citep[e.g.][]{secrestb2015,satyapal2016}, it is possible that the dust can be heated by star formation alone.  Since  there are a number of blue compact dwarfs (BCDs) with extremely red mid-infrared colors \citep[e.g.,][]{griffith2011,izotov2011}, it is possible that star formation in extreme and perhaps low metallicity environments can produce an infrared spectral energy distribution similar to an AGN. On the other hand, if the majority of bulgeless galaxies that display red mid-infrared colors are in fact AGNs, the fraction of AGNs in the low bulge mass regime has been significantly underestimated by optical studies. 
\par
The theoretical calculations presented in this work demonstrate that the extremely red $W1-W2$ mid-infrared colors observed in a few BCDs \citep[e.g.,][]{griffith2011,izotov2011} is not directly caused by low metallicity, contrary to what has been suggested by previous authors  \citep[e.g.,][]{hainline2016}. While low metallicity galaxies are linked to harder radiation fields  \citep[e.g.,][]{campbell1986,galliano2005,madden2006,hunt2010,levesque2010}, which in turn effects the strength of PAH features and the heating of the dust, the lower dust abundance requires very high column densities to produce to very red $W1-W2$ and $W2-W3$ WISE colors consistent with the few BCDs with anomalous colors reported in the literature(see Figure ~\ref{solarwisecolor}). 
Indeed, the vast majority of low metallicity galaxies do not have the extreme colors observed in the few BCDs with extreme colors reported in the literature \citep[e.g.,][]{griffith2011,izotov2011}, and most of the BCDs with anomalous colors do not have particularly low metallicities.  It should also be pointed out that, BCDs constitute only 6\% of the local dwarf population \citep{lee2009}, and many shows signs of pronounced kinematic and morphological disturbances   \citep[e.g.,][]{lelli2012a,lelli2012b} consistent with mergers \citep{bekki2008}. Given that dwarfs in the field are some of the most gas dominated galaxies in the local universe \citep{geha2006}, it is likely that the extremely red $W1-W2$ colors in the few BCDs reported in the literature are caused by a combination high ionization parameters and high column densities in gas rich low mass dwarfs with starbursts possibly triggered by mergers. This scenario would result in mid-infrared colors in the upper right corner of Figure ~\ref{solarwisecolor}, consistent with the infrared colors of the few anomalous BCDs reported in the literature \citep[e.g.,][]{griffith2011,izotov2011}.
\par
Our models show that optically classified starburst galaxies must be AGNs if they meet the theoretical AGN cut provided in this work.  Using our theoretical AGN cut, there are $\sim~3$ times as many mid-infrared selected AGN candidates in the dwarf galaxy sample from \citet{hainline2016} than is presented in Table 1 of their work. Note that while our models do not include the effects of shocks which can be produced from starburst-driven winds, such galaxies would not be characterized by an optical emission line spectrum typical of HII regions or star forming galaxies and location on the BPT diagram consistent with the star forming sequence \citep{kewley2001}. Therefore any galaxy with optical emission line ratios typical of star forming galaxies which meet our theoretical color cut {\it must} be an AGN. These will be promising  targets for future follow-up studies.

\section{Summary and Conclusions}

We have conducted for the first time a theoretical investigation of the mid-infrared spectral energy distribution (SED) produced by dust heated by an active galactic nucleus (AGN) and an extreme starburst. We used photoionization and stellar population synthesis models that extended past the ionization front in which both the line and emergent continuum is predicted from gas exposed to the ionizing radiation from a young starburst and an AGN. We note that the models we have presented here do not include the effects of shocks, which can affect both the standard optical spectroscopic diagnostics as well as the heating of the grains and the resulting transmitted mid-infrared SED. Our full suite of simulated SEDs are publicly available to download online.\footnote{\url{http://physics.gmu.edu/~satyapal/syntheticspectra}} Our main results can be summarized as follows:

\begin{enumerate}
\item{We find that an extreme starburst can mimic an AGN in single color mid-infrared color cuts employed in the literature using WISE. However, the starburst models with red $W1-W2$ colors comparable to AGNs are typically accompanied by redder $W2-W3$ colors. \\}

\item{We demonstrate that extreme starbursts can also mimic AGNs using three-band mid-infrared color-cuts employed in the literature for a narrow range of parameter space. However, the models that reproduce these colors require either extremely high ionization parameters or gas densities that are inconsistent with the observed optical emission line ratios of star forming galaxies.\\}

\item{We demonstrate that the low metallicity starburst models are not able to produce AGN-like mid-infrared SEDs.In instances of BCDs with extreme red mid-IR colors, our results show that these colors are likely due to a combination of high ionization parameters and large column densities.\\}

\end{enumerate}

 We have demonstrated that the mid-infrared colors are dependent on the incident radiation field and ISM conditions in complex ways.  In galaxies where both star formation and obscuration are significant, finding AGNs using mid-infrared color selection will be challenging. In such cases, infrared coronal line studies with JWST will offer an ideal opportunity to hunt for AGNs in optically star forming galaxies.  In a future paper, we will present the emission line spectrum predicted by these simulations and explore their diagnostic potential in finding elusive AGNs.

\section{Acknowledgements}

 We gratefully acknowledge the anonymous referee for their thorough review. We also thank Dillon Berger for his initial help in extracting infrared colors from our Cloudy runs.  This work greatly benefited from stimulating discussions with Moshe Elitzur, with whom it was a joy to discuss science. N.A.\ and S.S.\ gratefully acknowledge support by a Mason 4-VA Innovation grant.  N.J.S.\ held an NRC Research Associateship award at the Naval Research Laboratory for much of this work. We also gratefully acknowledge the use of the software \textsc{topcat} \citep{Taylor2005} and Astropy \citep{astropy2013}.\\

\end{document}